# Reconfigurable Multistate MRAM Synapses with Vortex STNO based Neurons for Scalable In-Memory Convolutional Neural Networks

**Ravish Kumar Raj[1,3,5], Simon N. Richter[3], Saeed Baghaee Ivriq[2], Oliver Fridorf[3], Darío Fernández-Khatiboun[1], Yasser Rezaeiyan[1], Luana Benetti[4], Tim Boehnert[4], Ricardo Ferreira[4], Hooman Farkhani[1], Sonal Shreya[3] and Farshad Moradi[1, 3]***

[1]The Faculty of Engineering, Institute of Mechanical and Electrical Engineering Department, IME, University of Southern Denmark, 5230, Odense

[2]Electronic Circuits and Systems, KU Leuven, Arenberg Campus, Kasteelpark Arenberg, Belgium

[3]Department of Electrical and Computer Engineering, Aarhus University, Aarhus, Aarhus N 8200, Denmark

[4]International Iberian Nanotechnology Laboratory (INL), Braga, Portugal

[5]Department of Electronics and Communication Engineering, Indian Institute of Technology, Roorkee, India 247667

*Email: moradi@sdu.dk

**Abstract:** Magnetic tunnel junction (MTJ)-based magnetic random-access memory (MRAM) is a promising platform for neuromorphic and in-memory computing owing to its non-volatility, high endurance, fast switching dynamics and CMOS compatibility. However, conventional spin-transfer torque and spin-orbit torque MRAM implementations for neural networks often suffer from high critical switching currents, large latency, thermal instability and significant read-write overheads. Here, we demonstrate a unified multistate MRAM-spin-torque nano-oscillator (STNO) architecture that integrates synapses and neurons on a single chip for convolutional neural network (CNN) applications. The system employs 1×8 multistate MRAM arrays as programmable synapses coupled with a vortex-based STNO neuron, enabling both individual and collective programming through field-line-driven write channels. Multiple configurable resistance states are achieved by tuning internal and external magnetic fields together with bias currents, allowing quantized positive and negative synaptic weights for configurable kernel and pooling operations. The proposed architecture is evaluated through simulation on MNIST, SVHN, CIFAR-10, Google Speech Commands (GSC) and RadioML datasets, achieving accuracy of 99.76%, 87.93%, 78.14%, 87.96% and 56.46% respectively. Based on fabricated device dimensions, the complete architecture occupies ~6171.2 µm² with an average energy consumption of 200.08 pJ per training and inference cycle for MNIST, highlighting its potential for scalable low-power neuromorphic computing.

## 1. Introduction

The rapid escalation of data-driven applications and artificial intelligence (AI) has revealed the limitations of traditional von Neumann architectures, which suffer from performance bottlenecks due to the physical separation of memory and computation units. This separation leads to high latency and significant energy consumption during frequent data transfers [1-3]. As conventional CMOS-based architectures approach their scaling and power limits, alternative paradigms inspired by biological neural systems have emerged to address these challenges [4-5]. Neuromorphic computing systems (NCSs), which emulate the architecture and functionality of the human brain, offer a promising route toward highly parallel, low-power, and adaptive information processing [6-7].

At the core of NCSs are neurons and synapses interconnected processing and memory units capable of executing as an in-memory computing task through event-driven communication mechanisms such as spikes or analog interactions [8-10]. Several technologies have been investigated for hardware realization of these units, including memristors [11-12], photonic devices [13-14], and spintronic elements [15-20]. Among them, spintronic-based components, particularly magnetic tunnel junctions (MTJs) based Magnetic Random Access Memory (MRAM), have garnered significant attention owing to their scalability, non-volatility, CMOS compatibility, and ultra-low energy requirements [21-29]. MTJ-based devices, as demonstrated in commercial MRAM technologies, offer an ideal platform for implementing artificial synapses and neurons with compact footprints and enhanced endurance [30-32].

Traditionally, binary MRAM devices utilizing parallel (P) and antiparallel (AP) magnetization states between the free and reference layer have been employed to emulate synaptic behavior. However, binary systems inherently limit synaptic resolution and, consequently, the learning capability of neuromorphic hardware [33-34]. To address this, recent research has introduced multistate MRAM architecture utilizing multiple MTJs based on spin-transfer torque (STT) and spin-orbit torque (SOT) mechanisms [15, 22]. These approaches allow finer modulation of synaptic weights, which can significantly enhance both learning accuracy and energy efficiency in neuromorphic computing [15,17,22]. However, multibit spintronic synapses, including domain wall (DW), magnetic skyrmion based and serially connected MTJ arrangements, have shown potential but often suffer from limited thermal stability and poor resistance contrast between states [15, 29, 35]. Furthermore, most existing parallel multistate MRAM configurations rely on a common read/write channel, which introduces latency, writing and reading overhead, and hampers the ability to dynamically reconfigure the device from multistate to single-state operation based on the input needs of NNs [15, 22]. This limitation restricts their applicability in adaptive synaptic designs with input nodes where reconfigurable weight granularity is essential for best performance across multiple benchmark datasets.

To address the limitations of existing binary and multistate MRAM-based synapses, we propose a reconfigurable multistate synaptic architecture based on a parallel 1×8 MRAM array integrated with a vortex-based STNO neuron on a single chip. Each MRAM cell in the array is coupled to a shared and individually addressable non-contact nanowire, referred to as an on-chip field line, which generates a localized Oersted field near the free layer (FL). This enables precise, energy-efficient, and reliable control of the FL magnetization state through a field-line-driven write mechanism. In addition, the device operation is further explored under externally applied magnetic fields generated via an equivalent toroidal coil configuration, enhancing tunability and programmability. The proposed architecture supports both individual and collective programming of MRAM cells, enabling robust and scalable realization of multilevel synaptic weights while significantly reducing read/write path overhead. In this configuration, each MRAM contributes discretely to the overall conductance, allowing the implementation of up to $N + 1$ distinguishable quantized weight levels for an array of $N$ MRAMs. The effective synaptic weights are further tuned by jointly modulating field-line (write) currents and scaled bias (read) currents through shared read/write channels. Leveraging the non-volatile and reconfigurable characteristics of MRAM, the architecture is well-suited for in-memory and hybrid analog-digital implementations of NNs. Using transistor-controlled MRAM arrays, configurable kernel and pooling matrices with quantized positive and negative weights are realized for CNN applications. The proposed design achieves multiple stable resistance states with high magnetoresistance contrast, enabling accurate weight representation. The effectiveness of the proposed system is validated through simulation across a range of benchmark datasets, including MNIST, SVHN, CIFAR-10, Google Speech Commands (GSC) (10 classes), and RadioML-2016, demonstrating competitive training accuracies within a quantized CNN framework. Furthermore, the architecture maintains reasonable area and power characteristics, highlighting its scalability and efficiency.

# 2. Results and discussion

## 2.1 MRAM characterization as synapse.

**Fig. 1** illustrates an overview of the hybrid MRAM-STNO array architecture for NCS. **Fig. 1 (a)** shows the microscopy image of fabricated device that comprises of 8 in-plane elliptical MTJ having dimension (0.1µm x0.5µm) and 1 STNO with circular dimension of 0.35µm x 0.35µm on single array having common and individual field line controlled by $V_{f1}$, $V_{f2}$, ..$V_{f10}$ on top (dotted blue box) for writing path and weighted inputs (bias) from top contact $v_{b1}, v_{b2} \ldots v_{b8}$ (dotted green box) to shared bottom contact (BC) pads for reading the states MRAM as a synapse. The MTJ cells are spaced 120 $\mu$m apart mostly due to the layout of the contacts pads and to eliminate stray field interactions between the neighboring MTJs. **Fig. 1 (b)** presents schematic of the MTJ stack that shows top field line and then bias line for input through top contact. Each top contact is connected with CoFeB/Ta/NiFe based free layer (FL) that is separated by MgO based tunnel barrier from reference layer (RL), where the RL magnetization is pinned by CoFeB based synthetic antiferromagnet (SAF) stack.

The complete material stack used in the devices investigated in this study consists of (thicknesses in nm [5 Ta / 25 CuN]x6 / 5 Ta / 5 Ru / 6 IrMn / 2.0 CoFe30 / 0.7 Ru / 2.6 $CoFe_{40}B_{20}$ / MgO / 2.0 $CoFe_{40}B_{20}$ / 0.21 Ta / 7

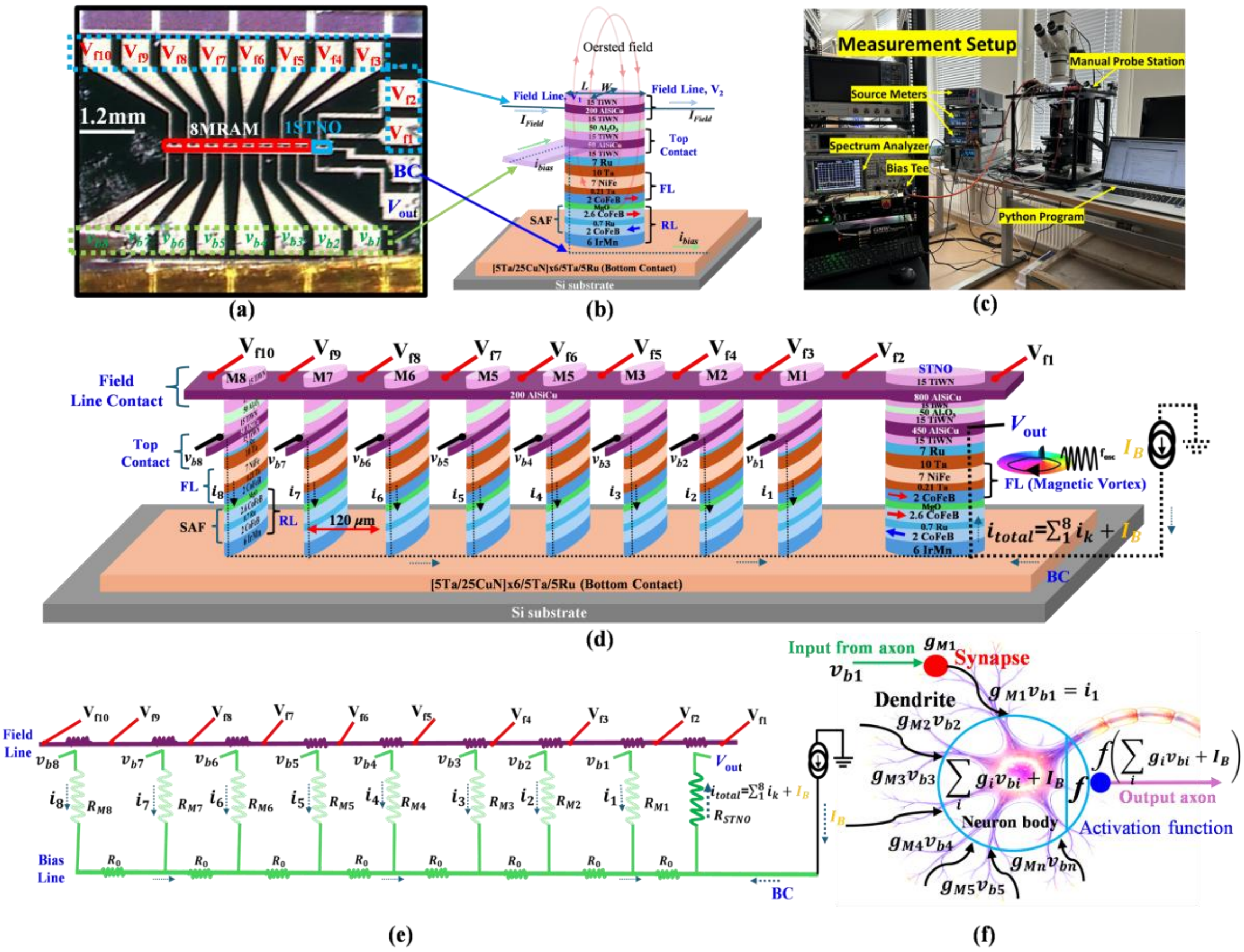

**Fig. 1**: Overview of the unified MRAM-STNO array architecture for neuromorphic computing system (a) Microscopic view of the fabricated device with 8 MRAM and 1 STNO having common and individual field line along with top and shared bottom contact pads (b) schematic of the MTJ stack (c) Measurement setup including source meters, spectrum analyzer, and probe station (d) Cross-sectional and array-level schematic illustrating the field-line configuration, top and shared bottom contact, and integrated STNO readout (e) Equivalent circuit representation showing field-line controlled by $V_{f1}$, $V_{f2}$, …..$V_{f10}$ and bias-line controlled by $v_{b1}(i_1)$, $v_{b2}(i_2)\ldots v_8(i_8)$ for routing the weighted STNO (f) Illustration of the neuron model, mapping MTJ currents to synaptic weights ( $g_M = \frac{1}{R_M}$ ) and activation function outputs from STNO characteristics.

NiFe / 10 Ta / 7 Ru. The MgO barrier was deposited with a wedge profile across the wafer, creating a controlled thickness gradient along the sample length and resulting in a resistance-area product (RA) ranging from 4 to 6 $\Omega\cdot\mu m^2$. The selected RA represents a trade-off between memory and oscillator performance, where higher RA values (~10 $\Omega\cdot\mu m^2$) are generally preferred for maximizing STNO oscillation power and frequency stability, while lower RA values are advantageous for MRAM operation due to reduced switching current and improved memory efficiency. The achieved intermediate RA range, combined with high TMR, therefore provides a favorable balance for the proposed reconfigurable MRAM-STNO architecture [16,17, 36-37]. At this MgO thickness, the nanodevices typically individual MRAM showed a TMR of 120 to 135% measured in a 4-terminals measurement such as 2 for field line and 2 for bias line. We investigate the switching characteristics of multiple MRAM and STNO by measuring magnetoresistance ($R$) through bias current ($i$) under sweeping internal field line current ($I$) and external field ($\boldsymbol{H}_{fl}$), and both for individual MRAMs and for a set of combined MRAMs connected via top electrodes and a shared common bottom contact. The detail characterization results of MRAM (see Table 1) and STNO device for single (1x8) and with various configurations are mentioned in Supplementary file **S1**.

**Fig. 1 (c)** shows the experimental setup used to characterize the device. The measurements are performed using a manual probe station, where electrical probes contact the device pads. The sample holder is used to mount the bared sample on probe station and two Keithly2450 source meter units are used to provide the sweeping field line current and bias current to the sample. The field line currents are used to make the transition to magnetization of FL while bias line current is used to measure the resistance of MRAM. At first, we apply a small current bias current ($i$ = 50$\mu$A) through the source meters to verify the probe connection with sample, if resistance value is observed in order of ohm, then measurement would be process further. The setup also includes a spectrum analyzer for detecting output frequency signals from the STNO, and a Bias-T to separate AC and DC components. The entire measurement process is automated using a Python-based control program connected to a computer. **Fig. 1 (d)** depicts the cross-sectional and array-level schematic illustrating the field-line configuration, top and shared BC, and integrated STNO readout. Individual MRAMs are excited by field lines and modulated the weighted current $i_1, i_2 \ldots i_8$ from the inputs $v_{b1}, v_{b2} \ldots v_{b8}$ as per its resistance states. The current flowing through each MRAM contributes to the total output current for the STNO. The cumulative current for STNO is expressed as:

$$I_{\text{total}} = \sum_{k=1}^{8} i_k + I_B \quad (1)$$

where $i_k$ is the current through each MRAM and $I_B$ represents the external bias current that is essential for STNO to be oscillation regime. The output voltage $V_{\text{out}}$ is measured across the STNO. Its equivalent circuit diagram of the 1 × 8 MRAM array used in measurements as depicted in **Fig. 1 (e)**. Each MRAM is modeled as a two terminal resistor whose value switches between $R_P$ and $R_{AP}$. The eight parallel resistors share a common BC node, and each top node is separately probed to extract cell by cell resistance, thereby isolating the intrinsic MRAM resistance from field line contact (write channel). **Fig. 1 (f)** illustrates the conceptual mapping of the MRAM array to a biological neuron model. Inputs from the axon terminals are weighted by synaptic conductance $g_M v_b$, representing synaptic weight $g_M = 1/R_M$. These signals propagate through the dendrites and accumulate in the neuron (STNO) body, where the weighted sum $\sum g_i v_{bi} + I_B$ is formed. The resulting signal passes through an activation function mapped to ReLU/Sigmoid/Softmax function with the characteristics of STNO, producing the output signal along the output axon. This representation demonstrates how the MRAM-STNO array can implement analog summation and activation functions, mimicking neural computation in hardware.

**Table 1: Overview of MRAM specifications**

| Devices | Cell dimension | TMR radio | Voltage/current threshold | Writing pulse width | Read-write overheads |
|---|---|---|---|---|---|
| MRAMs M1-M8 | 0.1μm x0.5μm | 120-130% | 0.8V/1mA | 1ns | No |

A key challenge in fabricating and controlling MRAMs states that it lies in achieving a bistable resistance state of elliptical MTJ stack on array sample with one STNO footprint of the nanopillar sized 0.35x0.35μm$^2$ that engineered for a vortex ground state [17, 37]. This is accomplished by introducing shape anisotropy through an elliptical device footprint. As a result, the MRAMs exhibit a largely square-shaped hysteresis loop with well-defined $R_P$ and $R_{AP}$ states at zero assisting field, along with two distinct coercive fields, as illustrated the characterization results under external and internal magnetic field with positive and negative bias current in **Fig. 2**. In the proposed device, the difference in coercive fields arises from device-to-device variation and MRAM position on BC from STNO, although it can be intentionally engineered by tuning the aspect ratio of the elliptical footprint.

The controlling state of MRAM by the internal field line such that a non-contacting nanowire is placed near to FL of each MTJ cell, generating an Oersted (Oe) magnetic field (see Supplementary file **S1**). The direction and magnitude of this field are controlled by the field line current ($I$) through source $V_{f1}$, $V_{f2}$ …... and $V_{f10}$ as shown in dotted light box in **Fig. 1** (a). The resulting Oersted field can be calculated using the Biot-Savart law, $\boldsymbol{H_{fl}} = \mu_0 I/4\pi r^3(\boldsymbol{dl} \times \boldsymbol{r})$ where $I$ is the applied field line current, $\boldsymbol{dl}$ is the differential length vector of the wire, and $\boldsymbol{r}$ is the position vector from the wire to the point of interest [35]. Given that the FL dimensions are significantly smaller than the length of the conducting nanowire, the wire can be approximated as infinitely long. Under this assumption, the equation simplifies to: $\boldsymbol{H_{fl}} = \mu_0 I/2\pi r$ [38]. This field directly influences the magnetization state of the MRAM's FL, which manifests as resistance changes in response to current recorded through resistance-vs-current R($I$) measurements. The R($I$) characteristics were obtained by sweeping the internal field line current ($I$) and external field ($\boldsymbol{H_{fl}}$) using one SMU, while a constant bias current was applied by second SMU, with the corresponding voltage across individual and multiple MRAM cells measured simultaneously. Hysteresis loops in *R(I)* were observed by sweeping the internal field line current from –140 mA to +140 mA. However, for external field an equivalent external field line current from –12mT to +12mT sweeping, using toroid (see Supplementary information **S1**). To investigate the free layer switching dynamics, a bias current ranging from –1 mA to +1 mA is applied along the top and bottom contacts of the device. The switching of the FL is driven by magnetic torque, proportional to ($\sim \boldsymbol{M} \times \boldsymbol{H_{fl}}$), where $\boldsymbol{M}$ denotes the effective magnetization of FL and $\boldsymbol{H_{fl}}$ is the effective magnetic field generated by the external field line current majorly and bias line current (minor effect by STT). **Fig. 2(a)** shows the magnetization hysteresis R(*I*) curve for MRAM1-MRAM8 using external field line sweep with bias currents of $i = 100\mu$A. The MR of each MRAM cell is defined as: $TMR = (R_{ap} - R_p)/R_p$. Distinct and symmetric switching P↔AP behavior is observed, with external field ($\boldsymbol{H_{fl}}$) of approximately at +5 to +7mT and at -8 to -11mT for various MRAMs. The observed switch between the resistance states $R_p$ (low-resistance) and $R_{ap}$ (high-resistance) corresponds to transitions of magnetization configurations of parallel and anti-parallel between FL and RL. **Fig. 2 (b)** shows the resistance switching of all MRAMs as a function of the internal field-line current sweep, which generates the local magnetic field near to FL and switched P↔AP at approximately +80 to +95mA and -120 to -135mA for various MRAMs under multiple sweeps. Slight variations among devices arise from fabrication tolerances and local magnetic variations, but the overall switching behavior remains consistent across the array.

The polarity dependence of the bias current clearly demonstrates the role of STT in modulating the switching process, including stepwise decoupling of the FL and RL. The near overlap of the R(*I*) hysteresis loops for polarity ±1 mA bias indicates that switching in MRAM is dominated by the damping-like STT, which symmetrically lowers the energy barrier for P↔AP transitions without introducing an effective bias field that would shift the loop along the current. **Fig. 2 (c)-(d)** presents the resistance switching of a representative MTJ measured under two opposite bias currents (+1.0 mA and −1.0 mA) as a function of the external magnetic field and internal field line. The nearly overlapping hysteresis loops indicate that the MTJ switching characteristics are largely symmetric with respect to bias polarity, confirming stable magnetic switching and minimal bias-induced distortion. The consistency of switching fields and resistance levels among the devices indicates uniform MTJ performance and fabrication quality. It is also noted that some MTJs exhibit non-symmetric or irregular switching behavior, likely due to fabrication-induced variability and thermal fluctuations.

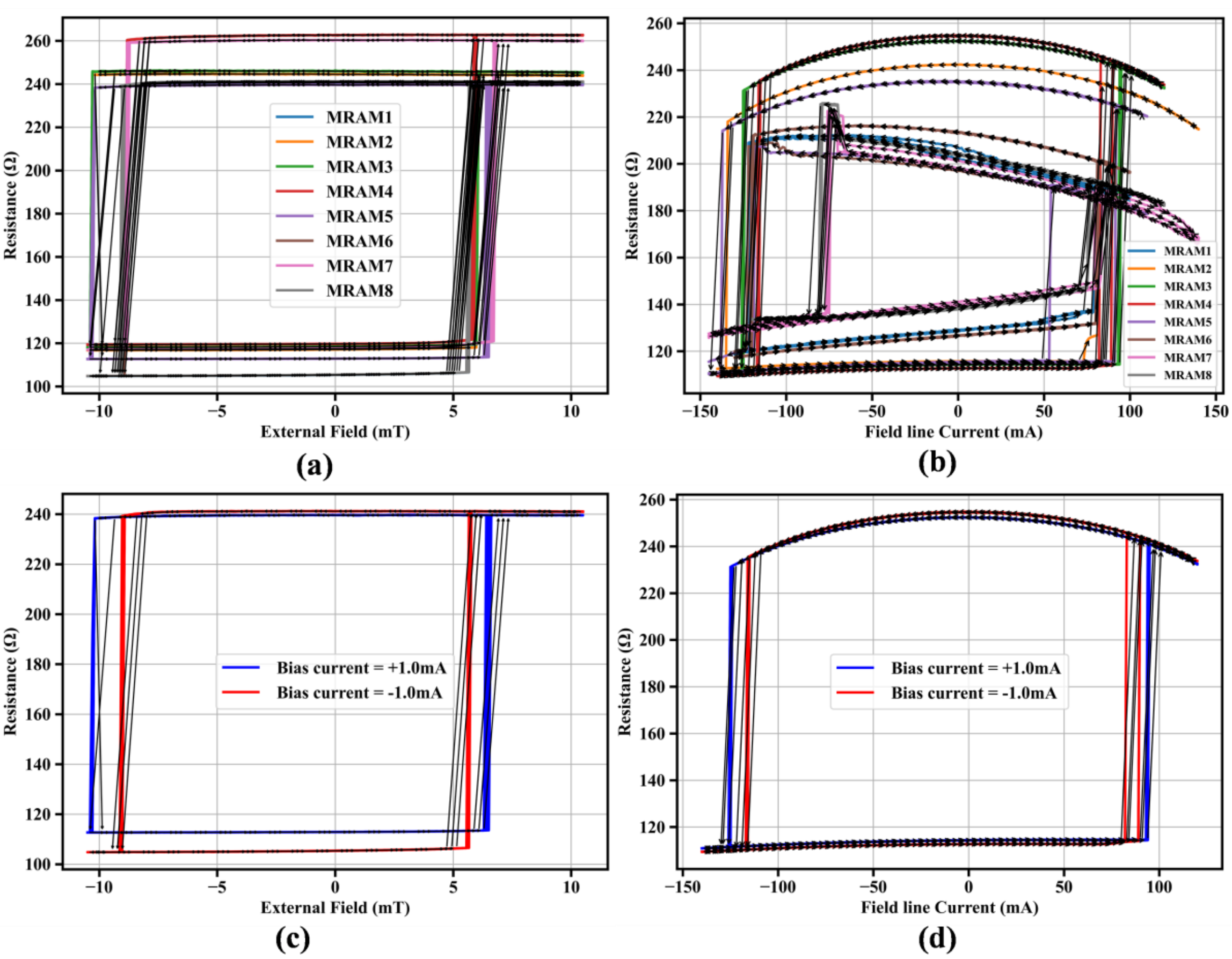

**Fig. 2**. Hystrsis of the fabricated MRAM array (a)-(b) show the resistance switching behavior of all eight (MRAM1-MRAM8) with multiple sweeps measured under external magnetic field and field-line current, respectively (c)-(d) present switching curves under positive and negative bias currents that is associted with positive and negative weights for synaptic connections.

**Fig. 3 (a)** shows the variation of measured resistance of each MRAM cell in both its parallel ($R_p$), and antiparallel ($R_{ap}$) states as a function of MRAM position (1 through 8) along the common bottom contact (BC). The variation shows the measured parallel resistance ($R_p$) and antiparallel resistance ($R_{ap}$) of individual MRAM cells as a function of their position relative to STNO (since STNO is closest to BC from right side in figure 1(d)). The $R_p$ slightly increases from approximately 104 Ω to 118 Ω, while the $R_{ap}$ increases from about 240 Ω to 250 Ω as the distance from the STNO increases due to the addition of 1-2 Ω path (line) resistance between two adjacent MRAM from BC. This gradual variation may originate from minor fabrication differences, line resistance, or position-dependent electrical pathways along the array. **Fig. 3 (b)** illustrates the equivalent resistance when multiple MRAM cells are electrically combined. As the number of MRAM devices increases from 1 to 8, the equivalent resistance decreases significantly for both magnetic states. The equivalent parallel resistance drops from roughly 110 Ω to about 15 Ω, while the equivalent antiparallel resistance decreases from around 240 Ω to nearly 30 Ω. This behavior follows the expected trend for resistors connected in parallel, where increasing the number of devices lowers the overall resistance. **Fig. 3 (c)** presents the magnetoresistance (MR) percentage as a function of the number of combined MRAM cells. The MR gradually decreases from about 131% to approximately 119% as more MRAM devices are connected. At the same time, the number of achievable resistance states (synaptic weights) increases linearly from 2 to 9 states with combined MRAM, enabling finer analog-like weight representation for neuromorphic computing. **Fig. 3 (d)** Shows the multi-level equivalent resistance states obtained from the combined MRAM configuration under sweeping external magnetic fields. The switching of individual MRAM cells produces multiple discrete resistance levels labeled I to IX (red mark), corresponding to different combinations of parallel and antiparallel states among the eight MRAM devices. When $N$ MRAM cells are connected in parallel, the resulting composite device exhibits a multilevel resistive behavior with $N$+1 distinguishable resistance states. This enables programming of the 8 MRAMs into 9 electrically distinguishable resistance state combinations of various P-P, P-AP, AP-P, and AP-AP state of respective MRAMs using short pulses delivered through independent field line contacts. The dashed lines indicate the intermediate resistance levels, while the solid curve represents the experimentally observed resistance transitions. These multiple stable resistance states enable the realization of multi-bit synaptic weights in spintronic neuromorphic architectures. Hence, combining multiple MRAM devices allows controllable equivalent resistance tuning, multi-level states, and increased synaptic weight resolution, which are key requirements for implementing hardware neural networks using spintronic devices.

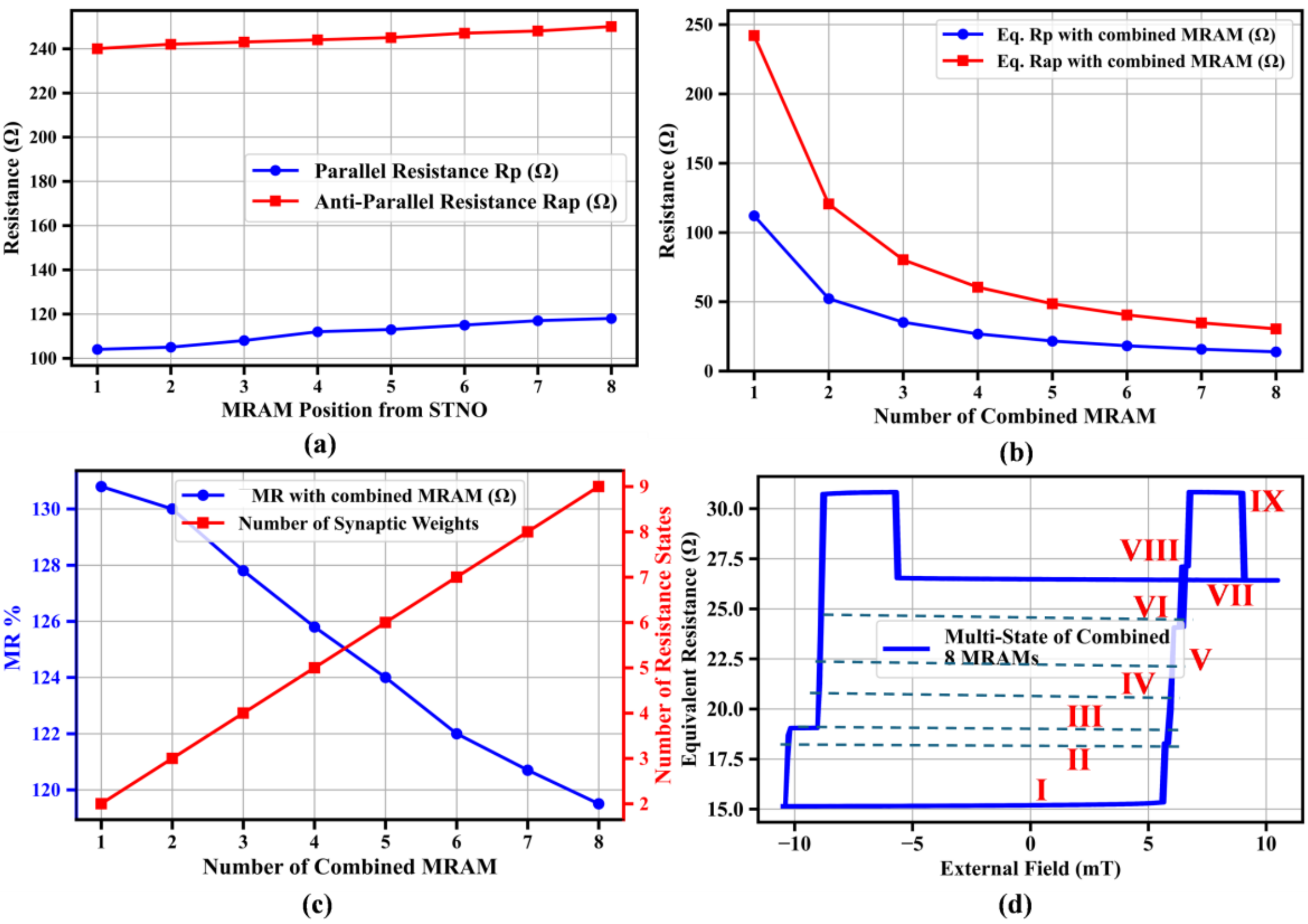

**Fig. 3**: Electrical characteristics and multi-state resistance behavior of the combined MRAM (a) $R_P$ and $R_{AP}$ variation with MRAM position from STNO on BC (b) Equivalent resistance of multiple combined MRAM scaling (c) MR evolution, and the resulting multi-level resistance states (d) Multilevel resistance states (I-IX) achieved through parallel combinations of 8 magnetic cells.

### 2.2. STNO characterization as neuron

As discussed in the previous section, the array comprises eight MRAM cells and one vortex-based STNO (see Table 2), which functions as a neuron. The STNO operates under a modulated bias current sourced (with input $v_{b1}$, $v_{b2}$ …$v_{b8}$) from each MRAM cell called as weighted input for STNO. But the only weighted current is not sufficient for circular motion of vortex in FL, therefore, a constant extra bias current $I_B$ (see as shown in **Fig. 1 (d)**) is required for stabilizing and circulating the vortex in FL. This device utilizes a circular MTJ stack with a vortex-configured FL to generate stable oscillations in the STNO resistance due to oscillation in net magnetization configuration between the vortex FL and RL. These oscillations yield output powers in the range of several µW under bias currents ranging from ±1 mA to ±6 mA, with field line current sweeps up to ±150 mA. During electrical characterization, a small bias current is initially applied to the STNO to ensure proper electrical contact with the SMU. Subsequently, the field line current is swept in the range of ±150 mA to examine the hysteresis behavior indicative of vortex formation at various bias currents, highlighted by light green strip (vortex state) as illustrated in **Fig. 4(a).** The R(*I*) characteristics of a vortex-based STNO measured under three different bias currents: -1.0 mA (blue) covered smallest area of loop, +1.0 mA (orange) and +0.5 mA (green) almost same area but shifted either side from each other due to opposing and supporting the switching at the critical points. Instead of abrupt binary switching, all curves exhibit a gradual, multi-step resistance evolution, which is a hallmark of vortex-state dynamics with grain where the vortex core progressively displaces and deforms under current [39-40]. The green shaded region highlights the vortex regime, in which resistance changes continuously due to vortex core motion rather than full magnetization reversal. The bias current shifts both the resistance level and the current range over which the vortex state is stable, demonstrating bias-dependent control of vortex nucleation, stability, and annihilation in the STNO. At a field line current of -140 mA, the free layer's magnetization aligns parallel to that of the reference layer. As the current increases to -40 mA the vortex is formed up to +120mA and finally magnetization reversed to an antiparallel state at +140 mA. Upon reversing the sweeping direction, the vortex state reemerges and remains stable in the range of -40 mA to -100 mA of field line currents.

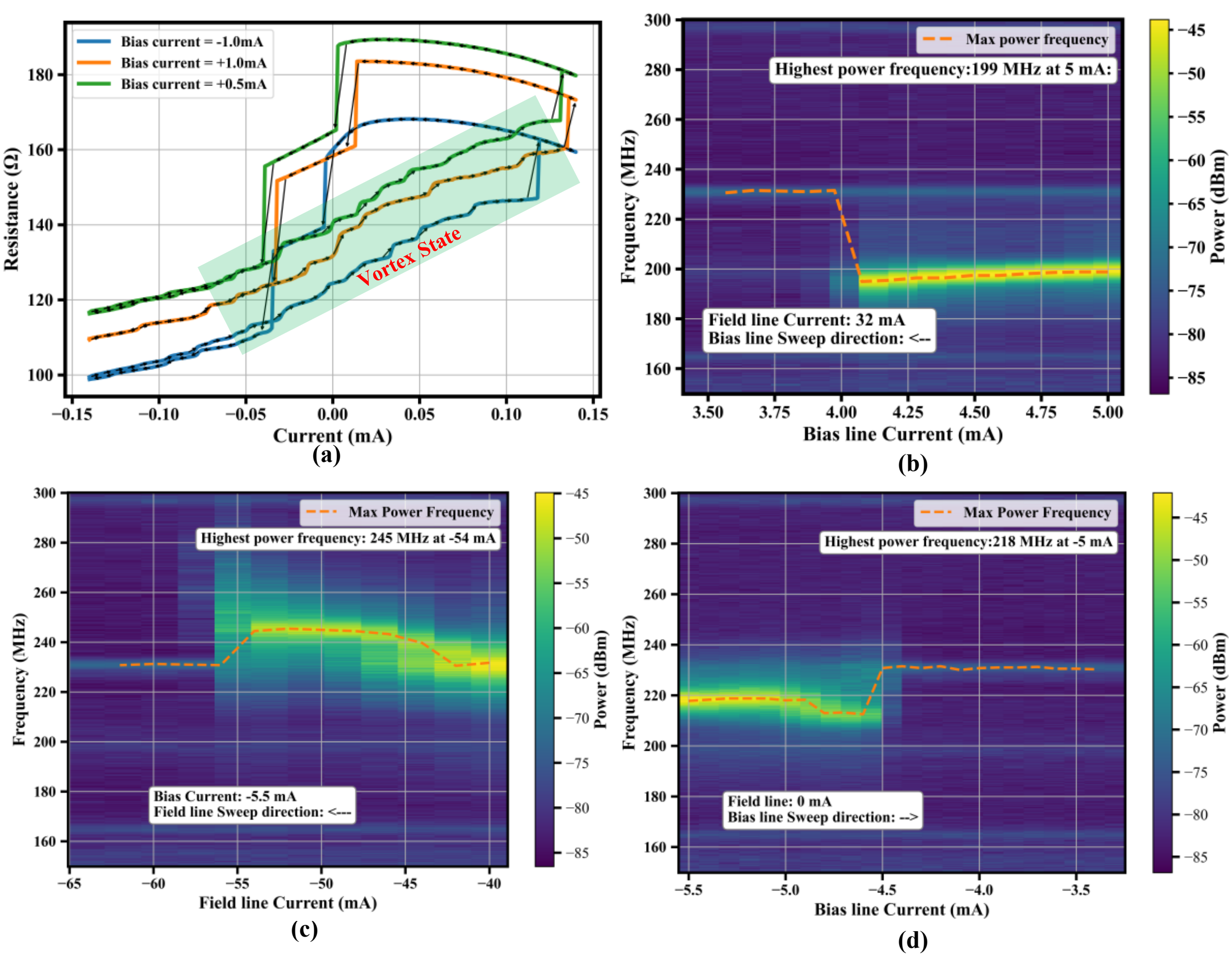

**Fig. 4:** Characterization results of oscillation (a) The hysteresis behavior indicates magnetization switching influenced by the bias direction and amplitude. and (b) Power spectral density corresponding to the bias current at a field line 32 mA bias. (c) Corresponding to the field line at bias current of 5.5 mA. (d) corresponding to the bias current at zero field line current.

**Table 2: Overview of STNO specifications**

| Devices | Cell dimension | TMR radio | Voltage/current threshold | Writing pulse width | Read-write overheads |
|---|---|---|---|---|---|
| STNO | 0.35µm x 0.35µm | 70-80% | 2V/5mA | 5ns | No |

The formation and stabilization of the vortex state arise from the balance between exchange and magnetostatic (dipolar) energies, which can be tuned using the field line and bias current [41-42]. Although the vortex state increases exchange energy due to non-uniform spin alignment, it greatly reduces stray field energy, making it energetically favorable [43]. The out-of-plane vortex core forms to maintain exchange continuity and avoid spin frustration. Its chirality and polarity can also be controlled by field line and bias currents [17, 36-37, 44]. Once formed, the vortex is excited into gyrotropic motion to generate resistance oscillations in the STNO. Under vortex conditions, the bias current becomes spin-polarized in the pinned layer and induces vortex dynamics in the free layer through spin-transfer torque (STT), Oersted fields, and localized heating [36-37]. When the current exceeds a critical threshold, the vortex core enters a sustained auto-oscillatory regime, where STT balances intrinsic magnetic damping [44]. This dynamic motion of the vortex core, relative to the in-plane magnetized RL, leads to resistance oscillations through the MR effect [17]. These oscillations enable electrical detection of the magnetic state and vortex position. The resulting output signal is analyzed using a Rohde & Schwarz spectrum analyzer (input resistance 50 Ω), connected via a Bias-T to isolate the AC component from the DC bias. The power spectral density (PSD) and frequency characteristics of the oscillations, under varying field line and bias current conditions with both polarities, are extracted using Lorentzian fitting of the measured spectra, as shown in **Fig. 4(b)-(d).**

The PSD exhibits a nonlinear threshold behavior, where the critical current represents the point at which intrinsic magnetic damping is completely offset by the STT effect. In the measurement, the STNO demonstrates a critical current of approximately 4 mA, with PSD increasing nearly linearly up to -40 dBm at 6 mA. The corresponding oscillation frequency is around 199 MHz. While vortex-based STNOs generally show an increase in frequency with bias current, the devices studied here display a nearly constant frequency response. This saturation behavior is attributed to operation near the upper limit of the frequency range, where the frequency becomes less sensitive to further increases in current**. Fig. 4(b)** presents the spectral response of the STNO for a bias current sweep from 4 mA to 5 mA under a constant field line current of 32 mA. The average oscillation frequency is 197 ± 2 MHz, with an output power of 35 ± 2.5 nW and a linewidth of 3.4 ± 0.5 MHz variations small enough to have a negligible impact on system performance. **Fig. 4(c)** illustrates that while frequency typically increases with field line current, saturation occurs at the upper frequency limit when the device is biased at -5.5 mA and field line current is -54 mA. Here, the average frequency reaches 238 ± 8 MHz, with an output power of 32 ± 2.5 nW and a linewidth of 3.5 ± 0.5 MHz again showing minimal variation. **Fig. 4(d)** depicts the response under zero field line current, with bias current swept from -5.5 mA to -3.5 mA. Even in the absence of field line input, the frequency saturates near the upper limit, yielding an average frequency of $215 \pm 3$ MHz, an output power of $30 \pm 2.5$ nW, and a linewidth of $5 \pm 0.5$ MHz. These stable parameters confirm reliable operation even without external magnetic field assistance. All measurements were conducted under an in-plane magnetic field generated by the field line current, without any out-of-plane component. This configuration is compatible with practical applications, as it can be easily realized using integrated local field sources in real-world spintronics devices.

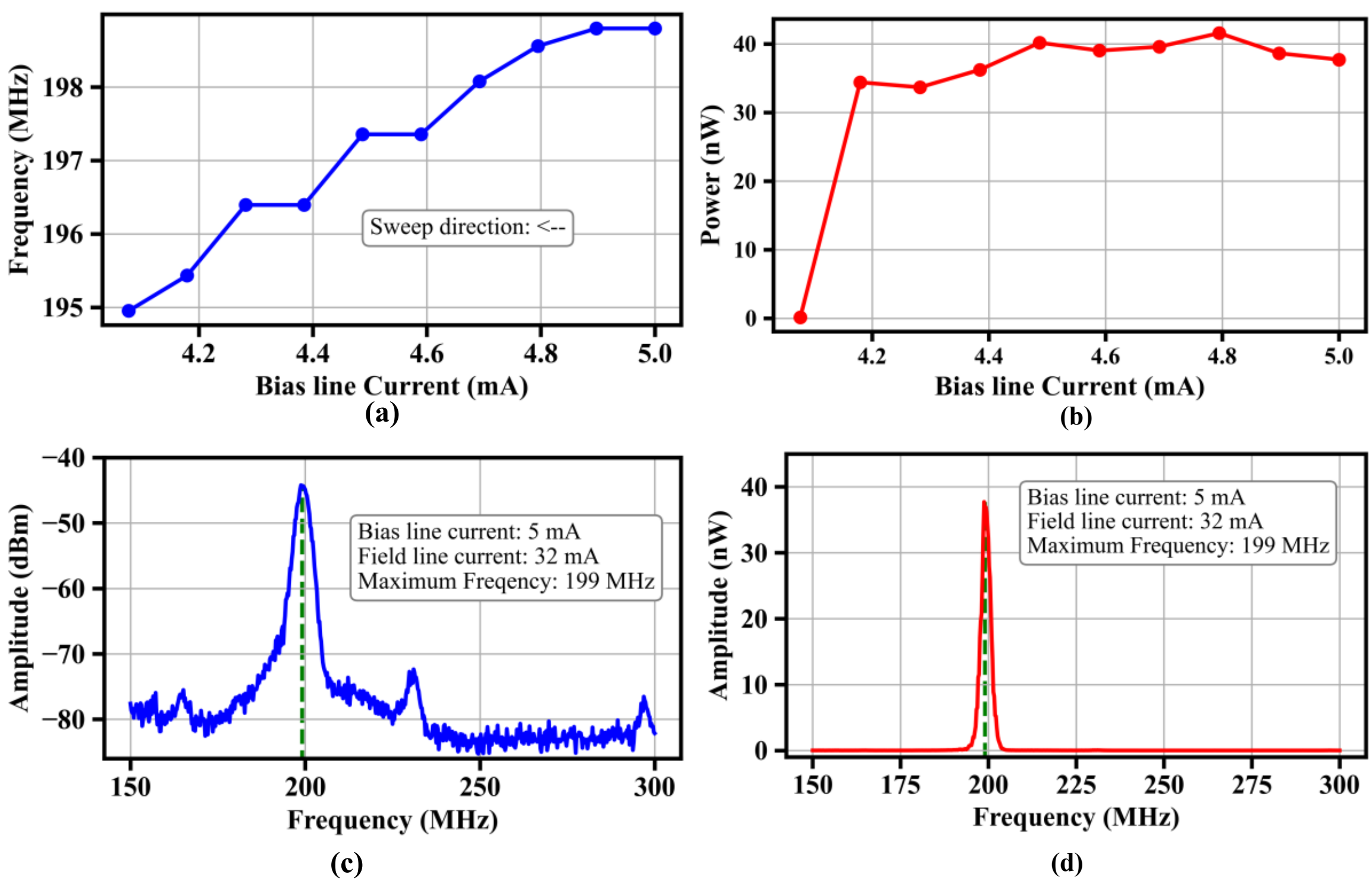

**Fig. 5:** Output Characteristics of STNO (a) shows the dependence of the oscillation frequency on the bias line current (b) Output power variation as a function of the bias current to the load (c) Power spectrum from frequency analyze (d) shows the processed spectral peak in amplitude, highlighting the oscillation signal centered at 199 MHz.

**Fig. 5(a)** shows the dependence of the oscillation frequency on the bias line current. As the bias current increases from approximately 4.1 mA to 5.0 mA, the oscillation frequency gradually increases from about 195 MHz to nearly 199 MHz. This monotonic increase indicates current-induced tuning of the STNO frequency through spin-transfer torque. The annotation indicates the sweep direction, showing that the current is swept in the reverse direction during the measurement. **Fig. 5(b)** presents the microwave output power as a function of the bias current to specific load. The oscillation power increases rapidly once the current exceeds the oscillation threshold, reaching values in the range of several µW that is suitable for integration with CMOS directly [17]. The power remains relatively stable over the current range with slight fluctuations, indicating a stable auto-oscillation regime of the STNO. **Fig. 5(c)** The bottom-left panel shows the measured microwave spectrum obtained from the spectrum analyzer. A strong spectral peak appears near 199 MHz, confirming the auto-oscillation frequency of the device. The peak amplitude reaches approximately -45 dBm, significantly above the background noise level. The measurement conditions indicated in the inset correspond to a bias line current of 5 mA and a field line current of 32 mA. **Fig. 5(d)** The bottom-right panel shows the processed spectral peak in voltage amplitude, highlighting the oscillation signal centered at 199 MHz. The peak amplitude reaches approximately µW, confirming the generation of a clear and stable microwave signal [17]. The inset summarizes the operating conditions: bias line current of 5 mA, field line current of 32 mA, and maximum oscillation frequency of 199 MHz. Hence, the current-tunable frequency generation, stable oscillation power, and well-defined spectral peak of the STNO, confirming its capability as a compact spintronic oscillator suitable for integrated neuromorphic computing system.

### 2.3 Configurable MRAM based kernel map and pooling layer with positive and negative weights for CNN

For efficient AI applications and optimal performance of neuromorphic devices and architectures, it is essential that the system design remains adaptive and flexible across different computational components [45-46]. These include the kernel and feature map matrices, pooling layers, as well as the input, hidden, and output layers, along with their corresponding synaptic weights and activation functions characteristics [47-48]. In this work, we introduce a configurable kernel mapping and pooling layer implementation scheme for hardware-based CNNs. The proposed approach enables the adjustment of speed and area efficiency and is experimentally validated using a fabricated 1×8 MRAM array. This framework provides flexible customization options that can accommodate diverse design requirements, thereby offering a scalable solution suitable for a wide range of applications. Such adaptability allows the device to be fine-tuned for different tasks and data characteristics. For example, consider a system designed to recognize numerical digits, whether handwritten or displayed on a computer screen. The configurations required for these two scenarios differ significantly. Handwritten digits often exhibit large variations in style, shape, and orientation depending on the writer, which necessitates more sophisticated processing and higher recognition precision. In contrast, digits displayed on a computer screen typically follow standardized fonts with minimal variation, requiring comparatively simpler processing. Consequently, there is growing interest in developing adaptive and reconfigurable neuromorphic systems capable of operating efficiently across diverse conditions and application requirements [49-53]. Such systems enhance the scalability, versatility, and overall efficiency of neuromorphic computing technologies. However, spintronic devices still face several challenges, including device-to-device variations, thermal stability issues, high write current requirements, endurance limitations, scalability of field-line currents, and the increasing impact of resistance in large-scale arrays.

#### 2.3.1 Design of configurable MRAM based kernel map and pooling layer matrix

The detailed configurable algorithm and its effectiveness is experimentally validated using the MRAM-STNO array platform. A key component of CNN-based neuromorphic computing is the implementation of the kernel and pooling layer mapping matrices, along with the fully connected artificial neural network (ANN), where synaptic weights and neurons serve as the fundamental processing units. **Fig. 6** illustrates a simplified schematic of a reconfigurable kernel mapping scheme using a single MRAM array, supporting kernel sizes of 2×4, 1×8, 2×2, 1×4, and 1×2, along with a 2×2 pooling layer for CNN architectures. For the 2×4 and 1×8 kernel configurations, all eight MRAM cells operate independently, providing four quantized weight levels (two positive and two negative). For the 2×2 and 1×4 configurations, pairs of MRAM cells are combined (2-2 configuration), enabling six quantized weight levels (three positive and three negative). Finally, for the 1×2 kernel configuration, four MRAM cells are grouped together (4-4 configuration), resulting in ten quantized weight levels (five positive and five negative).

The MRAM array is dynamically reconfigured to support multiple input-kernel configurations, specifically 8, 4, 2, and 1 input(s), forming a scalable input lattice as shown in the box of **Fig. 6** with blue color equal (=) sign meaning respective inputs are combined as single input. To achieve this flexibility, seven controlled transistors (T1-T7) are employed to control the bias-line (inputs) currents, selectively activating specific combinations of MRAM cells to realize the desired quantized synaptic weights. In our proposed design, individual bias and field lines for each MRAM cell can be selectively accessed via dedicated transistors (T1 through T7), allowing dynamic reconfiguration as illustrated by dark green color in **Fig. 6**. For eight-input architecture ($v_{b1}$-$v_{b8}$), all transistors remain OFF, enabling all eight inputs to function independently. In the 4-input mode, adjacent MRAM pairs are grouped via shared transistors (T1, T3, T5, T7), effectively halving the number of distinct inputs. Similarly, 2-input and 1-input modes are achieved by selectively activating larger MRAM groupings, controlled through specific transistor states for example, when only T4 is OFF, enabling full-array combination. This reconfigurability allows the system to dynamically adapt signal processing complexity, conserving power and optimizing performance in real time. Such scalable and energy-efficient architecture is particularly advantageous in neuromorphic systems, where the trade-off between input precision and computational load must be managed flexibly based on the task at hand.

**Table 3: Number of combined MTJ and resistance state in Ω as a quantized weight**

| #MMs \ States | R1 | R2 | R3 | R4 | R5 | R6 | R7 | R8 | R9 | R10 | R11 | R12 | R13 | R14 | R15 | R16 | R17 | R18 |
|---|---|---|---|---|---|---|---|---|---|---|---|---|---|---|---|---|---|---|
| 1 | 112 | 245 | -112 | -245 | | | | | | | | | | | | | | |
| 2 | 52.24 | 73.04 | 120.5 | -52.24 | -73.04 | -120.5 | | | | | | | | | | | | |
| 3 | 35.21 | 43.57 | 56.95 | 80.22 | -35.21 | -43.57 | -56.95 | -80.22 | | | | | | | | | | |
| 4 | 26.79 | 31.37 | 37.75 | 46.82 | 60.50 | -26.79 | -31.37 | -37.75 | -46.82 | -60.50 | | | | | | | | |
| 5 | 21.66 | 24.55 | 28.30 | 33.10 | 39.4 | 48.52 | -21.66 | -24.55 | -28.30 | -33.10 | -39.4 | -48.52 | | | | | | |
| 6 | 18.22 | 20.23 | 22.21 | 25.70 | 29.35 | 34.12 | 40.5 | -18.22 | -20.23 | -22.21 | -25.70 | -29.35 | -34.12 | -40.5 | | | | |
| 7 | 15.74 | 17.22 | 18.99 | 21.04 | 23.43 | 26.37 | 30.01 | 34.75 | -15.74 | -17.22 | -18.99 | -21.04 | -23.43 | -26.37 | -30.01 | -34.75 | | |
| 8 | 13.87 | 14.98 | 16.3 | 17.79 | 19.46 | 21.45 | 23.8 | 26.68 | 30.45 | -13.87 | -14.98 | -16.3 | -17.79 | -19.46 | -21.45 | -23.8 | -26.68 | -30.45 |

**Table 3** presents the quantized synaptic weight values obtained from the combined MRAM array for different kernel configurations and their corresponding weights. The first column indicates the number of combined MRAM cells (ranging from 1 to 8), while the remaining entries list the available quantized weight levels, including both positive and negative values. These signed weights enable excitatory and inhibitory synaptic operations within the neuromorphic architecture. For a single MRAM cell, the system supports four quantized weight levels, consisting of two positive values (112, 245) and two negative values (−112, −245), representing the most basic weight configuration. When four MRAM cells are combined, the number of available weights increases to ten quantized levels, including five positive values (26.79, 31.37, 37.75, 46.82, 60.05) and their corresponding negative counterparts (−26.79, −31.37, −37.75, −46.82, −60.05). Similarly, when eight MRAM cells are combined, the system achieves the highest weight resolution, providing eighteen quantized weight levels. These

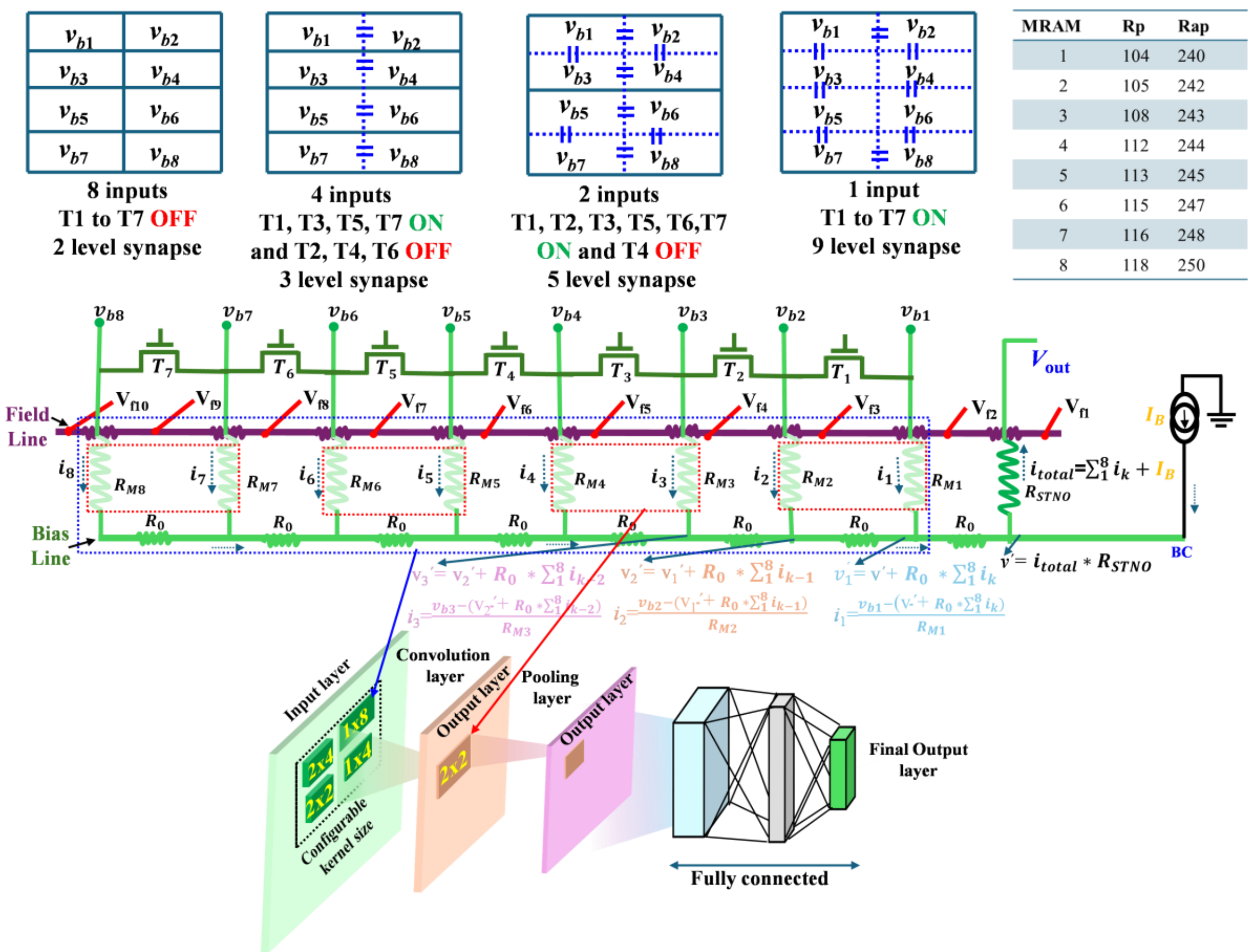

| MRAM | Rp | Rap |
|---|---|---|
| 1 | 104 | 240 |
| 2 | 105 | 242 |
| 3 | 108 | 243 |
| 4 | 112 | 244 |
| 5 | 113 | 245 |
| 6 | 115 | 247 |
| 7 | 116 | 248 |
| 8 | 118 | 250 |

**Fig. 6**: Schematic illustration of selective MRAM programming in a crossbar array, showing ON/OFF cell states, field and bias line operation, and the timing sequence for targeted device switching for configurable CNN architecture. Tabular form of resistance state of each MRAM.

include nine positive values (13.87, 14.98, 16.30, 17.79, 19.46, 21.45, 23.80, 26.68, 30.45) and their corresponding negative values (−13.87, −14.98, −16.30, −17.79, −19.46, −21.45, −23.80, −26.68, −30.45).

### 2.3.2 Design of positive and negative wights in the proposed array

The circuit in **Fig. 6** represents an actual controlling of the MRAM state using the field line and bias current through MRAM and external current source ($I_B$). The bias line on the top provides the input voltages $v_{b1}$to $v_{b8}$, while all BC cells are connected through small series resistors $R_0$ around 1-2 Ω, which is very small compared with the MRAM resistance (Rp~112 Ω and Rap~245). Each MRAM cell behaves like a resistor between the top contact of MRAM and BC. The current through the $n^{th}$ cell depends on the voltage difference between these two nodes. Because the BC of bias line contains the small series resistance $R_0$, the voltage along this line is not perfectly uniform; it changes depending on the sum of the currents flowing through all MRAM cells as shown the expression of node voltage with different color in **Fig 6**. Thus, the effective bias voltage at a particular node is influenced by the cumulative current flowing through the bias network. The current through the $n^{th}$ MRAM cell can therefore be written conceptually as:

$$i_n = \frac{v_{bn} - (\mathrm{v}_{n-1}' + R_0 \ * \sum_1^N i_{k-n-1})}{R_{M3}} > 0,\ if\ v_{bn} > (\mathrm{v}_{n-1} + R_0 \ * \sum_1^N i_{k-n-1})\ ;\text{for positive weight} \quad (2)$$

$$i_n = \frac{v_{bn} - (\mathrm{v}_{n-1}' + R_0 \ * \sum_1^N i_{k-n-1})}{R_{M3}} < 0,\ if\ v_{bn} < (\mathrm{v}_{n-1} + R_0 \ * \sum_1^N i_{k-n-1})\ ;\text{for negative weight} \quad (3)$$

Here, $V'$ and $v_{bn}$ are voltage at the BC and input voltage at respective MRAM. However, $R_0$ and $R_{Mn}$ are the line resistence of BC and MRAMs respectively. The external bias current source ($I_B$) at the rightmost side of the bias line plays a crucial role. It injects an external bias current into the bias network specially for bringing the STNO into oscillation regime and therefore supplies energy to the entire array along with input voltages $v_{b1}$-$v_{b8}$. Because of this external current source, the bias line can rise or fall in voltage depending on how the injected current is distributed through the MRAM branches and STNO resistance. If the voltage at the bias line of a particular cell $v_{bn}$ is higher than the BC voltage, current flows normally from the input through the MRAM device into the bias line. In this case the device behaves as a positive resistance (weight) element, meaning it dissipates power like a conventional resistor. However, due to the current supplied by the rightmost current source, the BC voltage at some nodes may become higher than the corresponding input voltage. When this happens the current through that MRAM branch reverses direction and flows to the input supply. In this situation the branch appears to exhibit negative resistance (weight), because the current direction is opposite to what would be expected from the applied input voltage alone. This does not mean the MRAM cell itself is physically a negative resistor (weight); rather, the energy provided by the right-side current source and the redistribution of current along the bias network causes the effective voltage difference to reverse, producing an apparent negative resistance (weight) condition [22,54].

In CNNs, positive kernel weights enhance relevant spatial features such as edges, gradients, and textures by increasing neuron activation, acting as excitatory synapses [55]. Negative weights suppress or differentiate features by reducing neuron activation, functioning as inhibitory synapses and enabling contrast or edge detection [56]. Similarly, in ANNs with fully connected layers, positive weights strengthen input influence while negative weights counteract it. The balance between excitatory and inhibitory weights allows neural networks to amplify important patterns, suppress irrelevant information, and effectively learn feature representations and decision boundaries [57]. Therefore, both CNN kernels and ANN fully connected layers rely on the combination of positive (excitatory) and negative (inhibitory) weights to properly learn feature representations and decision boundaries. This balance between excitation and inhibition allows the network to selectively amplify relevant patterns while suppressing irrelevant information, enabling efficient feature extraction and classification [58-60].

### 2.4 Configurable neuron functionality based on STNO characteristic

**Fig. 7** illustrates how the intrinsic nonlinear characteristics of the STNO can be utilized to realize configurable neuron activation functions for neuromorphic computing. By controlling the bias line current, the STNO output frequency and power can be tuned, enabling the implementation of different nonlinear activation behaviors commonly used in ANN. **Fig. 7 (a)** shows the normalized oscillation frequency of the STNO as a function of the normalized bias line current. The blue markers represent experimentally measured data, while the green dashed line represents a ReLU activation function fit with slope of 1.2. The frequency increases almost linearly once the bias current exceeds the oscillation threshold, closely matching the ReLU-like behavior. This demonstrates that the STNO can naturally implement a ReLU-type activation, where the output remains near zero below the threshold and increases linearly with increasing input current. **Fig. 7 (b)** presents the normalized output power of the STNO as a function of normalized bias current. The red markers represent the measured experimental data. The data fits with two common neural activation models: a sigmoid activation function (blue dashed curve) and a softmax-like activation behavior (green dashed curve). The gradual increase and saturation of the oscillation power closely resemble the nonlinear characteristics of these activation functions, showing that the STNO power output can emulate probabilistic or saturating nonlinear responses. These results demonstrate that the frequency and power outputs of the STNO can be configured to realize multiple activation functions, which are essential building blocks in neural network architectures [61]. The frequency response naturally resembles a ReLU function, while the power response exhibits sigmoid and softmax-like nonlinearities depending on the operating regime and normalization.

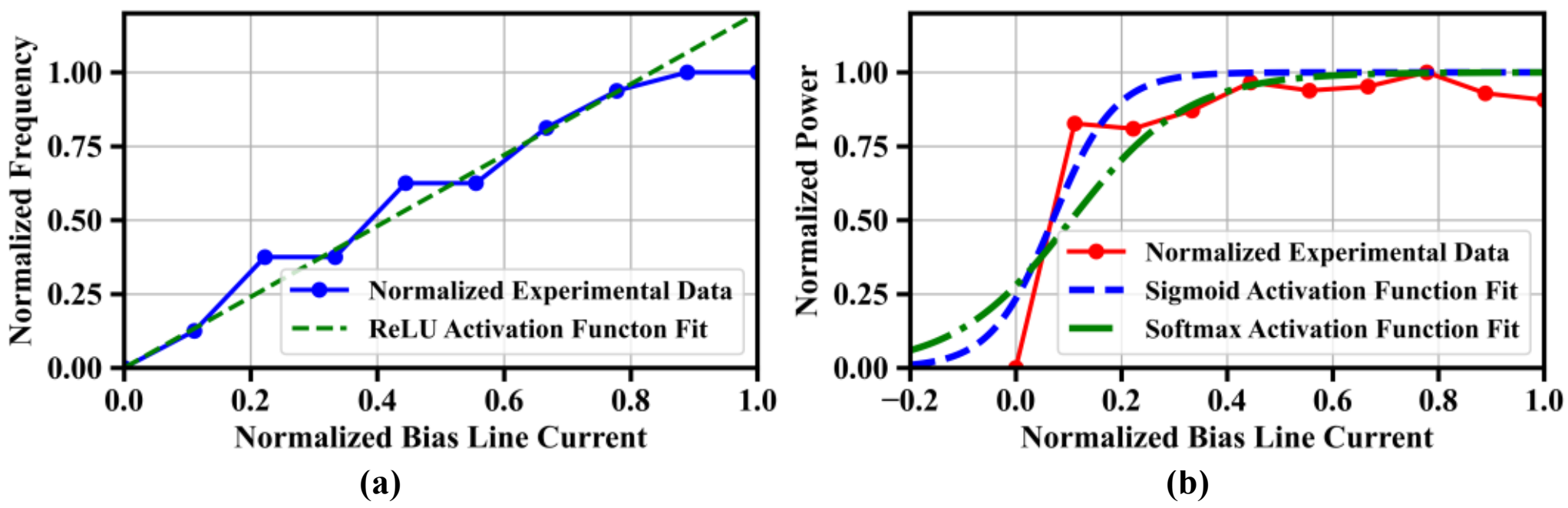

**Fig. 7:** Normalized frequency and power responses of STNO as a function of normalized bias line current (a) Demonstrating the capability of the STNO frequency to emulate ReLU neuron activation function (b) Power maps to sigmoid, and softmax neuron activation function.

## 3. CNN classification of digits for handwritten MNIST dataset

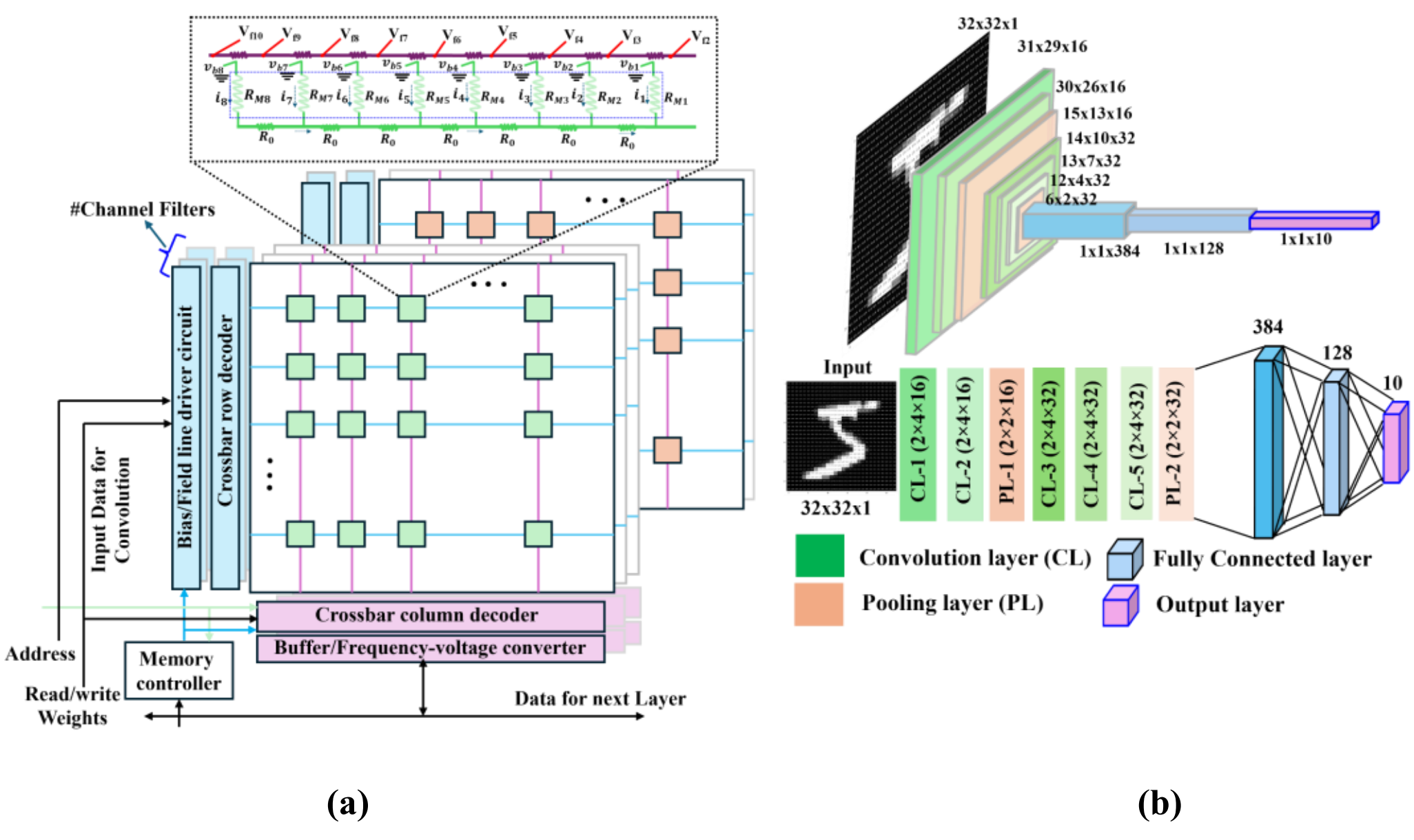

**Fig. 8**: (a) Hybrid configurable MRAM array organization for CNN application. (b) Illustration of the proposed MRAM-STNO based VGG-8 architecture for image classification on the MNIS dataset, consisting of convolutional layers (CL), pooling layers (PL), and fully connected (FC) layers.

To evaluate the performance of the proposed reconfigurable multistate MRAM-STNO neuromorphic architecture, CNN-based analyses are performed on benchmark datasets, including handwritten digit recognition tasks. The architecture employs multistate MRAM synaptic arrays and vortex-based STNO neurons to implement a quantized CNN, where stable MRAM resistance states represent positive and negative synaptic weights for convolution and pooling operations. Precise weight tuning is achieved through field-line-driven writing and bias-controlled reading. The system is first validated on the widely used MNIST handwritten digit dataset to assess classification accuracy and robustness [22, 55, 58-64]. To further demonstrate generalization capability of proposed structure, the system is extended for CNN analysis to more complex datasets such as SVHN, CIFAR-10, GSC (10 classes), and RadioML-2016 [65-72]. These datasets encompass a diverse range of input modalities, including grayscale images, color images, audio signals, and radio frequency data, thereby providing a comprehensive assessment of the proposed system.

**Fig. 8(a)** illustrates the operational flow of the proposed reconfigurable multistate MRAM-STNO crossbar architecture for CNN-based handwritten digit recognition. During convolution, the input feature map is scanned using programmable kernels of size 2×4 or 1×8, where the desired synaptic weights are written into selected MRAM cells through the write circuitry and crossbar row/column decoders. The dotted region represents the multistate MRAM synaptic array, in which each MRAM cell stores a quantized weight using its stable resistance states. The input image pixels are applied along the rows of the crossbar, while the programmed MRAM resistances modulate the resulting column currents, thereby performing parallel analog-like matrix-vector multiplication directly inside the memory array. The accumulated column currents are fed to vortex-based STNO neurons, which generate oscillatory/spiking responses according to the weighted summation. These oscillation frequencies are then converted into voltage signals using a frequency-to-voltage/analog to digital converter and forwarded to the next CNN layer [17, 73]. The architecture also includes peripheral circuits such as read/write drivers, field-line controllers, bias lines, and row/column decoders for accurate programming and selection of individual or multiple MRAM cells for multiple channel filters. The highlighted row configuration demonstrates how input voltages, together with field-line and bias currents, control the MRAM states and output currents. By integrating memory and computation within the same crossbar array, the proposed architecture enables highly parallel and energy-efficient in-memory CNN processing with reduced latency and read/write overhead compared to conventional digital implementations. **Fig. 8 (b)** presents the CNN pipeline used for classification of the MNIST handwritten digit dataset using VGG8 architecture [63, 74]. The input grayscale digit image is sequentially processed through multiple convolutional layers and pooling layers that extract hierarchical spatial features from the image [75]. Specifically, the network includes convolutional layers labeled CL1 to CL5, interleaved with pooling layers PL1 and PL2, which progressively reduce spatial resolution while increasing feature abstraction. The extracted feature maps are then flattened and passed through fully connected (FC) layers that perform the final classification into one of the ten-digit classes (0-9). The different colored blocks represent the successive transformations of feature maps as they propagate through the network.

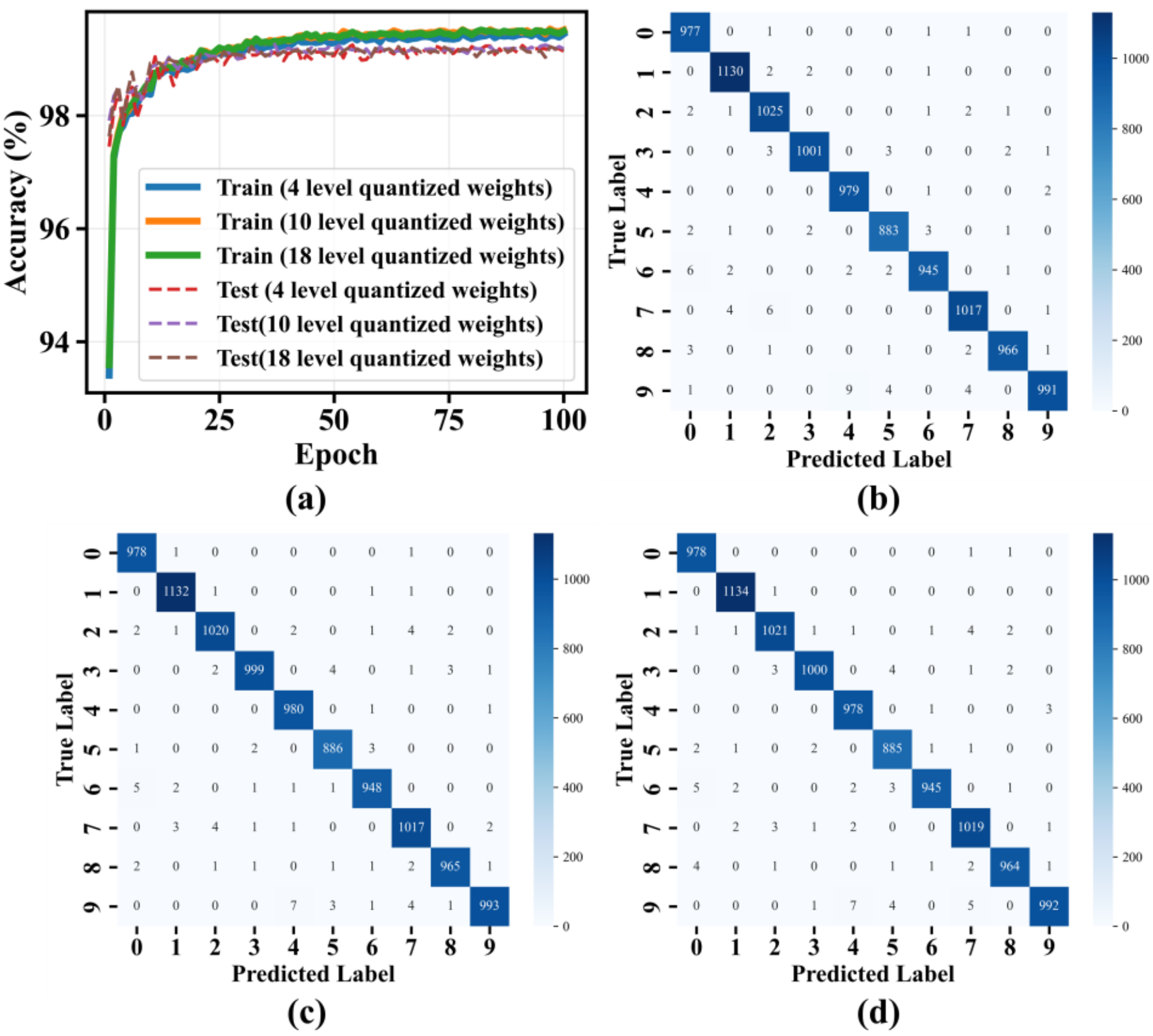

**Fig. 9:** (a) Shows the training accuracy progression over 100 epochs for 4-level, 10-level, and 18-level weight quantization. Heatmaps represent the confusion matrices of the trained models evaluated on the MNIST test dataset for (a) 4-level, (b) 10-level, and (c) 18-level quantization, illustrating the classification performance across digit classes 0-9.

**Fig. 9 (a)** illustrates the influence of weight quantization precision on the learning behavior and classification capability of the proposed convolutional neural network trained on the MNIST dataset. The evolution of training and testing accuracy over 100 epochs for networks employing 4-level, 10-level and 18-level weight quantization is achieved as 99.58%, 99.76% and 99.84% and 99.08%, 99.23% and 99.47%, respectively. Despite the strong constraint imposed by low-precision weights, all models successfully converge during training after 50 epoch. The network using 18 quantization levels exhibits the most stable learning behavior and achieves the highest accuracy, approaching ~99.84% toward the final epoch, reflecting the improved representational flexibility offered by a larger set of quantized weight states. **Fig. 9 (b)-(d)** present confusion matrices obtained from the evaluation of each trained model on the MNIST test dataset, providing a detailed view of the classification behavior across digit classes. In all cases, the matrices exhibit strong diagonal dominance, indicating that the majority of samples are correctly classified across all quantization settings. The 4-level quantized model already achieves high recognition accuracy for most digits, although some misclassifications appear among visually similar classes such as 3, 5 and 8. Increasing the number of quantization levels reduces these ambiguities, as observed in the 10-level configuration, where the diagonal elements become more pronounced and off-diagonal errors are reduced. The 18-level quantized model shows the most consistent classification performance, with minimal confusion between classes and the highest concentration of correct predictions along the diagonal. These results demonstrate that increasing quantization granularity improves both training stability and classification accuracy, while even highly constrained weight representations retain strong recognition capability. Collectively, the results highlight the effectiveness of quantization-aware training for enabling low-precision CNN implementations that maintain high classification performance, an important property for energy-efficient hardware accelerators and neuromorphic computing systems.

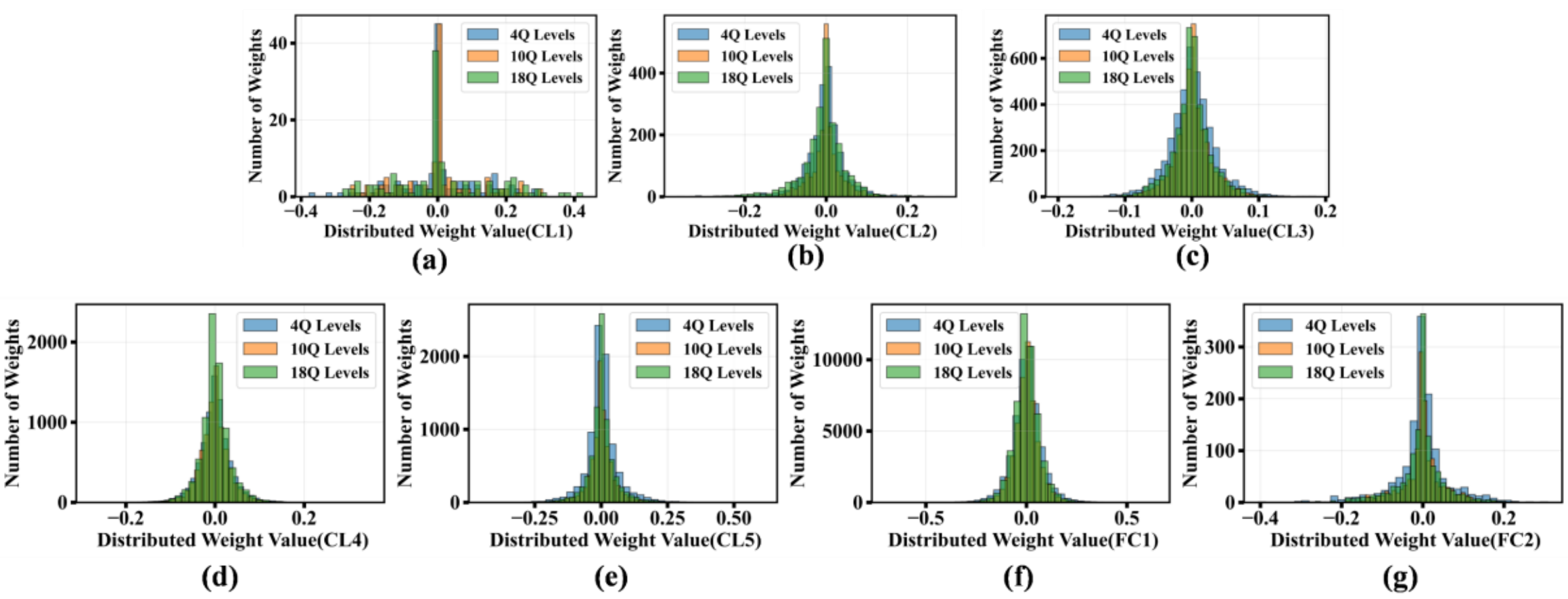

**Fig. 10 (a)-(g) Layer-wise comparison of weight distributions across different quantization levels in the CNN model.** Histograms are shown for all convolutional layers (CL1-CL5) and fully connected layers (FC1 and FC2), illustrates the learned weights are distributed after training and the quantization granularity influences the resulting weight representation.

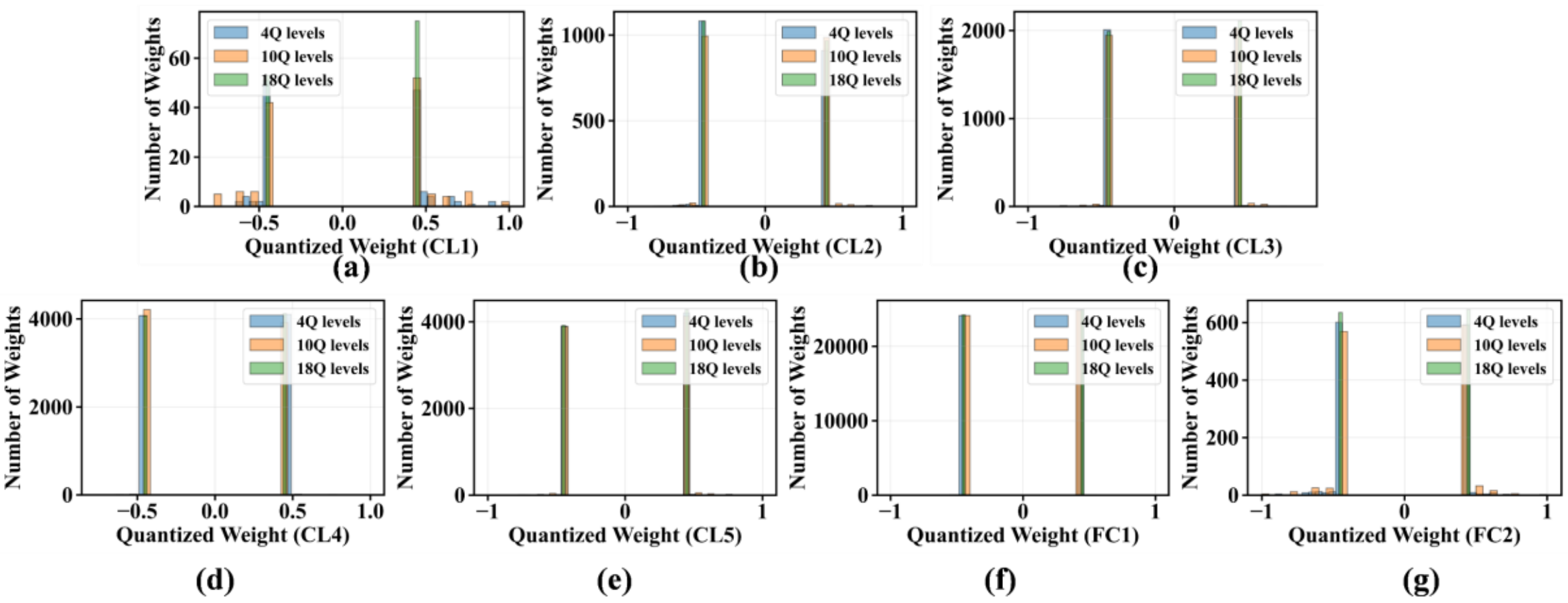

**Fig. 11 (a)-(g) Layer-wise comparison of quantized weight distributions across different quantization levels in the CNN model.** Histograms of quantized weights for all convolutional layers (CL1-CL5) and fully connected layers (FC1 and FC2)

**Fig. 10(a)-(g)** presents the weight distributions across various layers of the CNN implemented on the reconfigurable multistate MRAM-STNO architecture, illustrating the effect of different quantization levels on the mapped synaptic weights. Each histogram plots the number of weights against the distributed weight values for three quantization levels: 4Q (blue), 10Q (orange), and 18Q (green). The top row corresponds to the first three convolutional layers (CL1-CL3), where the weight distributions are tightly centered around zero, indicating that most synaptic weights are small in magnitude, while higher quantization levels (10Q and 18Q) provide finer resolution, capturing subtle variations in weights. The bottom row shows the weight distributions for deeper convolutional layers CL4-CL5 and the fully connected layers FC1-FC2, where the distributions are broader and the number of weights significantly increases, reflecting the accumulation of more synaptic connections in the deeper layers of the network. Across all layers, increasing the quantization levels results in smoother histograms and more accurately represented weight values, demonstrating the ability of the multistate MRAM array to encode fine-grained, discrete synaptic weights for CNN computations [15, 22]. Overall, the results of distribution highlights how the proposed architecture supports precise mapping of convolutional kernels and fully connected layers, maintaining high weight fidelity and enabling effective neural network performance with quantized synaptic representations.

The histogram shows in **Fig. 11 (a)-(g)** that in early convolutional layers (CL1-CL2), weights are more widely distributed across both negative and positive values, resulting in broader histograms with multiple peaks, and these layers are more sensitive to quantization, especially at lower levels like 4Q; as we move deeper into the network (CL3-CL5), the distributions become increasingly concentrated around a few dominant values, forming sharp spikes that indicate high redundancy and strong robustness to quantization, with minimal change across 4Q, 10Q, and 18Q levels; in the fully connected layers, FC1 exhibits an extremely tight, almost binary-like distribution with most weights clustered at a few values and very high counts, while FC2 shows slightly more spread but still maintains clear clustering, and overall, increasing quantization levels mainly refines the granularity of the distributions without significantly altering their shapes, especially in deeper layers.

## 4. Brief analysis of the proposed CNN architecture for SVHN, CIFAR-10, GSC, and RadioML datasets

**Fig. 12** presents a brief analysis of the proposed CNN architecture on the SVHN, CIFAR-10, GSC (10 classes), and RadioML-2016.10 datasets, highlighting dynamics, generalization capability, and stability of the network. The results provide insight into the trade-off between model precision and overall classification performance across different application domains. In **Fig. 12** (a) SVHN, the rapid rise and smooth saturation of both training and testing accuracy indicate that the task is relatively easy and the learned features are robust; the small gap between 4Q and higher quantization levels suggests that even coarse quantization preserves most of the essential information, while slightly higher accuracy for 10Q/18Q reflects improved representation fidelity. In **Fig. 12** (b) CIFAR-10, the slower convergence and larger fluctuations in test accuracy highlight increased task complexity and sensitivity to weight precision; here, finer quantization (10Q and 18Q) better captures subtle feature variations, leading to modest performance gains and smoother optimization, while 4Q introduces quantization noise that slightly degrades generalization. For audio processing, a 10-keyword subset of the Google Speech Commands dataset was employed, containing short utterances of the words "yes," "no," "up," "down," "left," "right," "on," "off," "stop," and "go"; each waveform was transformed into a Mel-frequency cepstral coefficient representation with 40 coefficients over 128 time frames, resulting in a 5120-dimensional feature vector that was standardized using the training-set mean and variance [76]. In **Fig. 12** (c) Speech Commands, the pronounced oscillations in both training and testing accuracy indicate instability during learning, which can be physically interpreted as the model struggling to settle into an optimal parameter configuration due to both dataset variability and quantization-induced discretization errors; higher quantization levels reduce this effect by allowing more precise weight updates, thereby improving stability and final accuracy. For RF modulation classification, the RadioML 2016.10a dataset was used, comprising complex baseband I/Q sequences corresponding to three modulation formats (BPSK, QPSK, and 16-QAM); only samples with signal-to-noise ratio SNR $\geq$ 18 dB were considered, and for each example the in-phase and quadrature components of length 128 were converted to magnitude and phase, concatenated into a 256-dimensional standardized feature vector [77]. In **Fig. 12** (d) RadioML 2016, the nearly overlapping curves for all quantization levels and the quick saturation at lower accuracy suggest that the task is either inherently limited by feature separability or noise, making it less sensitive to weight precision; thus, even aggressive quantization (4Q) retains sufficient representational capacity. Overall, the graphs show that quantization acts as a form of constraint or noise on the weight space: lower levels (4Q) impose stronger discretization, which can reduce model capacity and introduce optimization instability, while higher levels (10Q and 18Q) approximate full-precision behavior more closely, improving convergence smoothness and accuracy; however, the impact depends strongly on dataset complexity, with simpler or more redundant tasks being more robust to aggressive quantization. **Table 4** compares different neural network architectures, including VGG and ResNet models, evaluated on benchmark datasets such as MNIST, SVHN, CIFAR-10, Speech Command, and RadioML. The table summarizes important performance and hardware metrics, including network size, average classification accuracy, estimated hardware area of the proposed CNN architecture, and average power consumption per read/write operation. These results highlight the trade-offs between classification performance, architectural complexity, area overhead, and energy efficiency. The classification accuracy values are obtained through simulations performed for the respective datasets, architectures, and network configurations. In contrast, the approximately estimated area and average power consumption are derived from the experimentally fabricated single-array MRAM-STNO sample with dimensions of $1.2$ mm $\times$ $0.5$ $\mu$m, (see Fig.1(a)) consisting of 8 MRAM cells ($0.5$ $\mu$m $\times$ $0.1$ $\mu$m) and 1 STNO device ($0.35$ $\mu$m $\times$ $0.35$ $\mu$m). The hardware estimation corresponds to the proposed neural network architecture excluding CMOS peripheral and ADC/FVC footprint. Energy consumption is calculated using operating conditions of $V_{\text{read}} = 0.6$ Vand $V_{\text{write}} = 1.8$ V, with a write pulse width of 5 ns applied during device characterization using source meters.

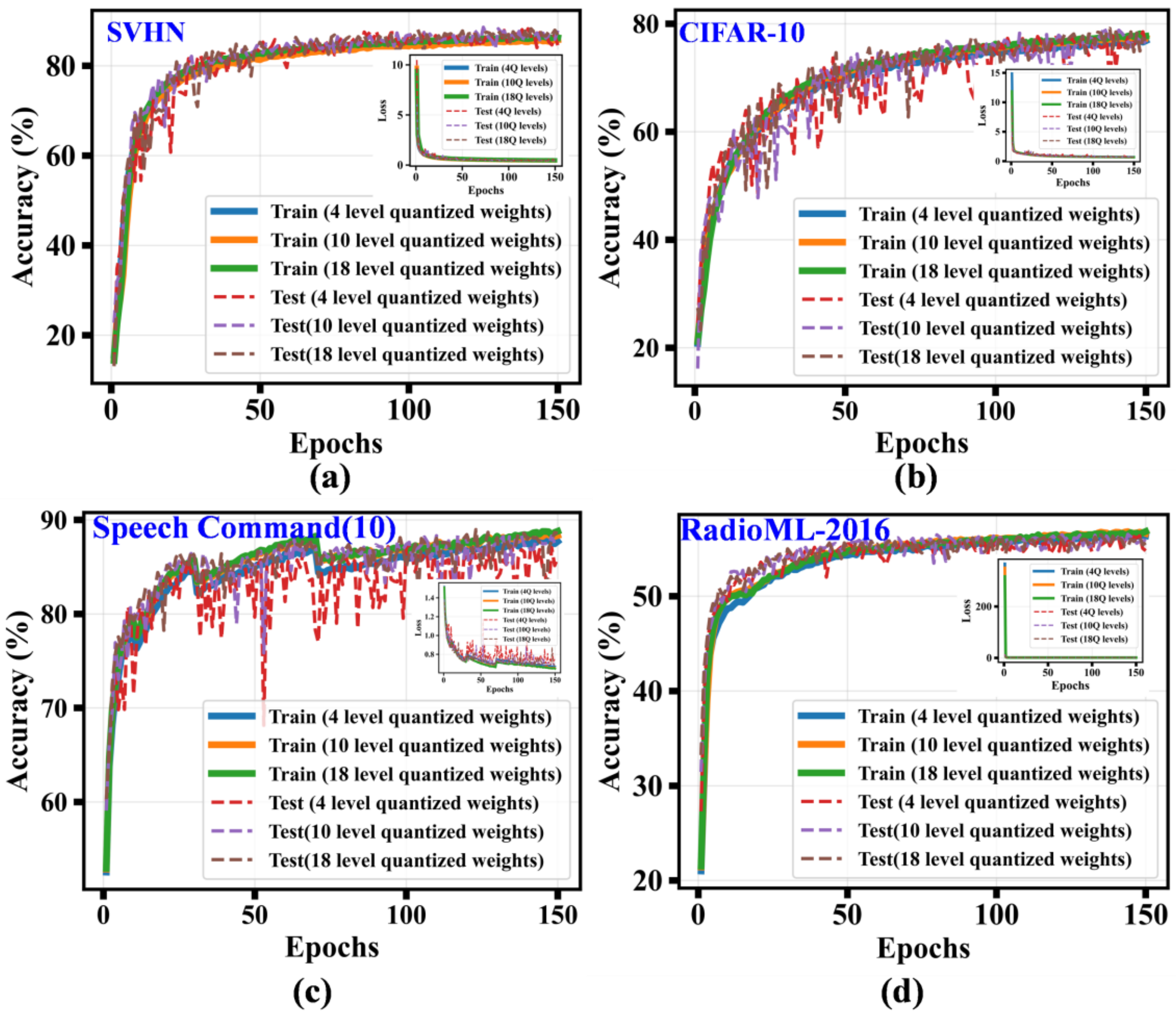

**Fig. 12** Training and testing accuracy versus epochs for neural networks with different weight quantization levels (4, 10, and 18 levels) across four datasets: (a) SVHN, (b) CIFAR-10, (c) Speech Commands (10 classes), and (d) RadioML 2016. Insets show the corresponding training and testing loss curves.

**Table 4: Performance Metrics of Neural Network Architectures Across Benchmark Datasets**

| Datasets | Architecture | Network size | Average Accuracy | Estimated Area on MRAM&STNO | Average Energy (read+write)/operation |
|---|---|---|---|---|---|
| MNIST | VGG8 | 5CL, 2PL, 2FC | 99.76% | 6171.2μm$^2$ ~78.5μm×78.5μm | 80fJ+200pJ=200.08pJ |
| SVHN | VGG16 | 6CL, 3PL, 2FC | 87.93% | 6818.5μm$^2$ ~82.5μm×82.5μm | 115fJ+280pJ=280.11pJ |
| CIFAR-10 | VGG16 | 6CL, 3PL, 2FC | 78.14% | 6818.5μm$^2$ ~82.5μm×82.5μm | 115fJ+280pJ=280.11pJ |
| Speech Command (10) | ResNet | 3CL, 1PL, 2FC | 87.96% | 4355.8μm$^2$ ~66.0μm× 66.0μm | 38fJ+106pJ=106.04pJ |
| RadioML 2016 | ResNet | 3CL, 0PL, 2FC | 56.46% | 4225.3μm$^2$ ~65.0μm× 65.0μm | 35fJ+98pJ=98.04pJ |

## 5. Conclusion

In conclusion, we have demonstrated reconfigurable multistate MRAM-STNO neuromorphic architecture for efficient in-memory implementation of convolutional neural networks, addressing fundamental limitations of the memory-compute separation in conventional AI hardware. The proposed framework integrates 1×8 MRAM arrays as synaptic elements with vortex-based STNOs as neurons on a unified chip, enabling compact, scalable, and energy-efficient neural computation directly within the memory substrate. The multistate MRAM design supports multiple stable resistance levels within a single device, allowing accurate encoding of quantized positive and negative synaptic weights and thereby enhancing computational density without increasing circuit complexity. The proposed device operates with internal field line sweeping currents (-140 mA to 140 mA) and external magnetic field sweeps (-12 mT to 12 mT) as the write channels, while a bias (weighted) current (-1 mA to +1 mA) is employed as the read channel for MRAM resistance. The resulting MR response captures diverse MRAM state combinations, counts, and dissimilarities, governed by the effective coupling of shared reading and writing pathways. This enables highly configurable synaptic encoding and reliable discrimination of resistance states under varying operating conditions. The field-line-driven programming scheme, combining internal and external magnetic field control with bias-current modulation, supports both individual and collective weight updates, thereby reducing read/write overhead and improving energy efficiency relative to conventional STT- and SOT-based MRAM implementations.

The architecture has been validated through simulation across multiple benchmark datasets, including MNIST, SVHN, CIFAR-10, GSC (10 classes), and RadioML-2016, demonstrating robust and consistent performance across image and signal-processing tasks within a quantized CNN framework. These results confirm the adaptability and generalization capability of the proposed system for heterogeneous AI workloads. System-level analysis further indicates a compact footprint of ~6171.2 µm² excluding CMOS and ADC/FVC peripheral and an energy consumption of only 200.08 pJ per training and inference operation for MNIST, underscoring its suitability for ultra-low-power edge and embedded intelligence applications. Moreover, synaptic weight distribution analysis reveals that higher quantization levels facilitate smoother kernel mapping, improved convergence stability, and enhanced learning performance. Overall, the proposed multistate MRAM-STNO architecture provides a promising route toward scalable, energy-efficient, and high-performance neuromorphic computing, offering a strong foundation for next-generation neural accelerators and hybrid analog-digital AI hardware systems.

## 6. Methods and Device Fabrication

The spintronic devices (MRAM) used in this work are called MTJ and consist of the bottom electrode, the MTJ stack with a pinned layer, a tunnel barrier, and a free layer, and the capping structure including the top electrode. The material stack is deposited by magnetron sputtering in a Timaris MTM and through patterning using e-beam lithography and ion-beam milling, a nanopillar is formed. Subsequently, the bottom contact is microstructure using laser beam lithography, the sample is covered in 800 nm $SiO_2$ and planarized by ion-beam milling under various angles to open an electrical contact to the top of the nanopillar. The bottom contact material stack consists of [5 Ta / 25 CuN]x6 / 5 Ta / 5 Ru (thickness in nanometers). The MTJ stack contains 6 IrMn / 2.0 CoFe30 / 0.7 Ru / 2.6 $CoFe_{40}B_{20}$ / MgO / 2.0 $CoFe_{40}B_{20}$ / 0.21 Ta / 7 NiFe / 10 Ta / 7 Ru. The MgO layer serves as the tunneling barrier and the MgO layer thickness defines the resistance-area (R×A) product of the MTJ. The 6 IrMn / 2.0 $CoFe_{30}$ / 0.7 Ru /2.6 $CoFe_{40}B_{20}$ layers are known as synthetic antiferromagnetic, which acts as the pinned layer, and the 2.0 CoFe40B20 / 0.21 Ta / 7 NiFe layer are the free layer. The top contact is deposited together with a second thick electrode on top called field line, which is later used to apply a local Oersted field to the MTJ. This top electrode stack consists of 15 TiWN / 50 AlSiCu / 15 TiWN / 50 $Al_2O_3$ / 15 TiWN / 200 AlSiCu / 15 TiWN, which is subsequently micro structured in two steps. First the field line and then the bottom electrode. Finally, the whole wafer is magnetic annealed for two hours at a magnetic field of 1 T to align the easy axis of the free layer along a defined in-plane direction of the wafer.

### Data availability:

The data that support the findings of this study are available from the corresponding author upon reasonable request. The Python scripts used for data visualization, basic fitting, and neuromorphic simulations are available from the corresponding author upon reasonable request.

### Acknowledgements

This project was fully supported by funding from the European Union's Horizon 2020 research and innovation program under grant agreement No. 899559 of SpinAge

### Competing interests

The authors declare no competing interests.

### Additional information

One supplementary information

#### Author Contributions

H.F., F.M., and R.F. have visualized the project and constructed the research framework. R.K.R, S.R., S.B.I., O.F., D.F. K. Y.R., and S.S performed the measurement of the devices, carried out the CNN simulation and drafted the manuscript. T.B., L.B., and R.F. designed and fabricated the samples. All authors reviewed and commented on the manuscript.

### References

[1] Shulaker, M.M., Hills, G., Park, R.S., Howe, R.T., Saraswat, K., Wong, H.S.P. and Mitra, S., 2017. Three-dimensional integration of nanotechnologies for computing and data storage on a single chip. *Nature*, *547*(7661), pp.74-78.

[2] Yu, S., 2018. Neuro-inspired computing with emerging nonvolatile memorys. *Proceedings of the IEEE*, *106*(2), pp.260-285.

[3] Markov, I.L., 2014. Limits on fundamental limits to computation. *Nature*, *512*(7513), pp.147-154.

[4] Poon, C.S. and Zhou, K., 2011. Neuromorphic silicon neurons and large-scale neural networks: challenges and opportunities. *Frontiers in neuroscience*, *5*, p.108.

[5] Sourikopoulos, I., Hedayat, S., Loyez, C., Danneville, F., Hoel, V., Mercier, E. and Cappy, A., 2017. A 4-fJ/spike artificial neuron in 65 nm CMOS technology. *Frontiers in neuroscience*, *11*, p.123.

[6] Dong, Z., Lai, C.S., Zhang, Z., Qi, D., Gao, M. and Duan, S., 2021. Neuromorphic extreme learning machines with bimodal memristive synapses. *Neurocomputing*, *453*, pp.38-49.

[7] Li, Y. and Ang, K.W., 2021. Hardware implementation of neuromorphic computing using large-scale memristor crossbar arrays. *Advanced Intelligent Systems*, *3*(1), p.2000137.

[8] Mehonic, A., Sebastian, A., Rajendran, B., Simeone, O., Vasilaki, E. and Kenyon, A.J., 2020. Memristors From in-memory computing, deep learning acceleration, and spiking neural networks to the future of neuromorphic and bio-inspired computing. *Advanced Intelligent Systems*, *2*(11), p.2000085.

[9] Shahsavari, M., Thomas, D., van Gerven, M., Brown, A. and Luk, W., 2023. Advancements in spiking neural network communication and synchronization techniques for event-driven neuromorphic systems. *Array*, *20*, p.100323.

[10] Yao, P., Wu, H., Gao, B., Tang, J., Zhang, Q., Zhang, W., Yang, J.J. and Qian, H., 2020. Fully hardware-implemented memristor convolutional neural network. *Nature*, *577*(7792), pp.641-646.

[11] Saïghi, S., Mayr, C.G., Serrano-Gotarredona, T., Schmidt, H., Lecerf, G., Tomas, J., Grollier, J., Boyn, S., Vincent, A.F., Querlioz, D. and La Barbera, S., 2015. Plasticity in memristive devices for spiking neural networks. *Frontiers in neuroscience*, *9*, p.51.

[12] Zhou, G., Wang, Z., Sun, B., Zhou, F., Sun, L., Zhao, H., Hu, X., Peng, X., Yan, J., Wang, H. and Wang, W., 2022. Volatile and nonvolatile memristive devices for neuromorphic computing. *Advanced Electronic Materials*, *8*(7), p.2101127.

[13] Shastri, B.J., Tait, A.N., Ferreira de Lima, T., Pernice, W.H., Bhaskaran, H., Wright, C.D. and Prucnal, P.R., 2021. Photonics for artificial intelligence and neuromorphic computing. *Nature Photonics*, *15*(2), pp.102-114.

[14] Li, R., Gong, Y., Huang, H., Zhou, Y., Mao, S., Wei, Z. and Zhang, Z., 2025. Photonics for Neuromorphic Computing: Fundamentals, Devices, and Opportunities. *Advanced Materials*, *37*(2), p.2312825.

[15] Rzeszut, P., Chęciński, J., Brzozowski, I., Ziętek, S., Skowroński, W. and Stobiecki, T., 2022. Multi-state MRAM cells for hardware neuromorphic computing. *Scientific reports*, *12*(1), p.7178.

[16] Costa, J.D., Serrano-Guisan, S., Lacoste, B. *et al.* High power and low critical current density spin transfer torque nano-oscillators using MgO barriers with intermediate thickness. *Sci Rep* **7**, 7237 (2017).

[17] Böhnert, T., Rezaeiyan, Y., Claro, M.S., Benetti, L., Jenkins, A.S., Farkhani, H., Moradi, F. and Ferreira, R., 2023. Weighted spin torque nano-oscillator system for neuromorphic computing. *Communications Engineering*, *2*(1), p.65.

[18] Martins, L., Jenkins, A.S., Alvarez, L.S.E. *et al.* Non-volatile artificial synapse based on a vortex nano-oscillator. *Sci Rep* **11**, 16094 (2021).

[19] Soni, S., Raj, R.K., Das, K., Rezaeiyan, Y., Boehnert, T., Farkhani, H., Ferreira, R., Moradi, F. and Shreya, S., 2025, December. Temperature-Robust Frequency-Encoded Spintronic Oscillatory Neural Networks for Energy-Efficient Edge AI. In *2025 First International Conference on Intelligent Computing and Systems at the Edge (ICEdge)* (Vol. 1, pp. 1-6). IEEE.

[20] Raj, R.K., Bindal, N., Poonia, V.S., Puliafito, V., Finocchio, G., Moradi, F. and Shreya, S., 2025, July. SkyNeu: Energy Efficient Antiferromagnetic Skyrmion Based Artificial Neuron for Neuromorphic Computing. In *2025 IEEE 25th International Conference on Nanotechnology (NANO)* (pp. 334-339). IEEE.

[21] Zahedinejad, M., Fulara, H., Khymyn, R., Houshang, A., Dvornik, M., Fukami, S., Kanai, S., Ohno, H. and Åkerman, J., 2022. Memristive control of mutual spin Hall nano-oscillator synchronization for neuromorphic computing. *Nature materials*, *21*(1), pp.81-87.

[22] Kumar, A., Lin, D.J., Das, D., Huang, L., Yap, S.L., Tan, H.R., Tan, H.K., Lim, R.J., Toh, Y.T., Chen, S. and Lim, S.T., 2024. Multistate Compound Magnetic Tunnel Junction Synapses for Digital Recognition. *ACS Applied Materials & Interfaces*, *16*(8), pp.10335-10343.

[23] Hong, J., Stone, M., Navarrete, B., Luongo, K., Zheng, Q., Yuan, Z., Xia, K., Xu, N., Bokor, J., You, L. and Khizroev, S., 2018. 3D multilevel spin transfer torque devices. *Applied Physics Letters*, *112*(11).

[24] Raymenants, E., Vaysset, A., Wan, D., Manfrini, M., Zografos, O., Bultynck, O., Doevenspeck, J., Heyns, M., Radu, I.P. and Devolder, T., 2018. Chain of magnetic tunnel junctions as a spintronic memristor. *Journal of Applied Physics*, *124*(15).

[25] Daddinounou, S. and Vatajelu, E.I., 2022, April. Synaptic control for hardware implementation of spike timing dependent plasticity. In *2022 25th International Symposium on Design and Diagnostics of Electronic Circuits and Systems (DDECS)* (pp. 106-111). IEEE.

[26] Soni, S., Rezaeiyan, Y., Boehnert, T., Farkhani, H., Ferreira, R., Kaushik, B.K., Moradi, F. and Shreya, S., 2025. SpinONN: energy efficient brain-inspired spintronics-based Hopfield oscillatory neural network for image denoising. *Neuromorphic Computing and Engineering*, *5*(3), p.034001.

[27] Hong, J., Li, X., Xu, N., Chen, H., Cabrini, S., Khizroev, S., Bokor, J. and You, L., 2020. A Dual Magnetic Tunnel Junction-Based Neuromorphic Device. *Advanced intelligent systems*, *2*(12), p.2000143.

[28] Rajib, M.M., Bindal, N., Raj, R.K., Kaushik, B.K. and Atulasimha, J., 2024. Skyrmion-mediated nonvolatile ternary memory. *Scientific Reports*, *14*(1), p.17199.

[29] Siddiqui, S.A., Dutta, S., Tang, A., Liu, L., Ross, C.A. and Baldo, M.A., 2019. Magnetic domain wall based synaptic and activation function generator for neuromorphic accelerators. *Nano letters*, *20*(2), pp.1033-1040.

[30] Jung, S., Lee, H., Myung, S., Kim, H., Yoon, S.K., Kwon, S.W., Ju, Y., Kim, M., Yi, W., Han, S. and Kwon, B., 2022. A crossbar array of magnetoresistive memory devices for in-memory computing. *Nature*, *601*(7892), pp.211-216.

[31] Ikegawa, S., Mancoff, F.B., Janesky, J. and Aggarwal, S., 2020. Magnetoresistive random access memory: Present and future. *IEEE Transactions on Electron Devices*, *67*(4), pp.1407-1419.

[32] Kwon, S., Myung, S., An, J., Kim, H., Kim, M., Lee, H., Yi, W., Jung, S., Yoon, D., Han, S. and Chung, S., 2025, February. 37.3 Monolithic in-Memory Computing Microprocessor for End-to-End DNN Inferencing in MRAM-Embedded 28nm CMOS

Technology with 1.1 Mb Weight Storage. In *2025 IEEE International Solid-State Circuits Conference (ISSCC)* (Vol. 68, pp. 1-3). IEEE.

[33] Sengupta, A. and Roy, K., 2016. Short-term plasticity and long-term potentiation in magnetic tunnel junctions: Towards volatile synapses. *Physical Review Applied*, *5*(2), p.024012.

[34] Prasad, N., Pramanik, T., Banerjee, S.K. and Register, L.F., 2020. Realizing both short-and long-term memory within a single magnetic tunnel junction based synapse. *Journal of Applied Physics*, *127*(9).

[35] Qiu, S., Zeng, J., Han, X. and Liu, J., 2024. On-chip skyrmion synapse regulated by Oersted field. *AIP Advances*, *14*(3).

[36] Oberbauer, F., Winkel, T.J., Böhnert, T., Wanjura, C.C., Claro, M.S., Benetti, L., Çaha, I., Deepak, F.L., Moradi, F., Ferreira, R. and Münzenberg, M., 2025. Magnetic tunnel junctions driven by hybrid optical-electrical signals as a flexible neuromorphic computing platform. *Communications Physics*, *8*(1), p.329.

[37] Li, R., Rezaeiyan, Y., Böhnert, T., Schulman, A., Ferreira, R., Farkhani, H. and Moradi, F., 2024. Temperature effect on a weighted vortex spin-torque nano-oscillator for neuromorphic computing. *Scientific Reports*, *14*(1), p.10043.

[38] Tawfik, S.A., 2025. A microscopic theory for the calculation of magnetic field induction of nanowires. *Computational Materials Science*, *248*, p.113611.

[39] Fridorf, O., Bjørnskov, L.M.R., Jenkins, A., Benetti, L., Shreya, S., Rezaeiyan, Y., Böhnert, T., Ferreira, R., Moradi, F. and Farkhani, H., 2025, May. Granular Spintronics-based Reservoir Computing for Temporal Applications. In *2025 IEEE International Symposium on Circuits and Systems (ISCAS)* (pp. 1-5). IEEE.

[40] Shreya, S., Jenkins, A.S., Rezaeiyan, Y., Li, R., Böhnert, T., Benetti, L., Ferreira, R., Moradi, F. and Farkhani, H., 2023. Granular vortex spin-torque nano oscillator for reservoir computing. *Scientific Reports*, *13*(1), p.16722.

[41] Pribiag, V.S., Krivorotov, I.N., Fuchs, G.D., Braganca, P.M., Ozatay, O., Sankey, J.C., Ralph, D.C. and Buhrman, R.A., 2007. Magnetic vortex oscillator driven by dc spin-polarized current. *Nature physics*, *3*(7), pp.498-503.

[42] Dussaux, A., Khvalkovskiy, A.V., Bortolotti, P., Grollier, J., Cros, V. and Fert, A., 2012. Field dependence of spin-transfer-induced vortex dynamics in the nonlinear regime. *Physical Review B—Condensed Matter and Materials Physics*, *86*(1), p.014402.

[43] Metlov, K.L. and Guslienko, K.Y., 2002. Stability of magnetic vortex in soft magnetic nano-sized circular cylinder. *Journal of magnetism and magnetic materials*, *242*, pp.1015-1017.

[44] Shreya, S., Rezaeiyan, Y., Jenkins, A., Böhnert, T., Farkhani, H., Ferreira, R. and Moradi, F., 2022. Verilog-A-based analytical modeling of vortex spin-torque nano oscillator. *IEEE Transactions on Electron Devices*, *69*(8), pp.4651-4658.

[45] Rajendran, B., Simeone, O. and Al-Hashimi, B., 2026. Towards efficient and reliable artificial intelligence through neuromorphic principles. *Philosophical Transactions of the Royal Society A: Mathematical, Physical and Engineering Sciences*, *384*(2315).

[46] Malviya, R.K., Danda, R.R., Maguluri, K.K. and Kumar, B.V., 2024. Neuromorphic computing: Advancing energy-efficient AI systems through brain-inspired architectures. *Nanotechnology Perceptions*, *20*(14), pp.1548-1564.

[47] Li, B., Qi, P., Liu, B., Di, S., Liu, J., Pei, J., Yi, J. and Zhou, B., 2023. Trustworthy AI: From principles to practices. *ACM Computing Surveys*, *55*(9), pp.1-46.

[48] Ivanov, D., Chezhegov, A., Kiselev, M., Grunin, A. and Larionov, D., 2022. Neuromorphic artificial intelligence systems. *Frontiers in Neuroscience*, *16*, p.959626.

[49] Xu, M., Chen, X., Guo, Y., Wang, Y., Qiu, D., Du, X., Cui, Y., Wang, X. and Xiong, J., 2023. Reconfigurable neuromorphic computing: materials, devices, and integration. *Advanced materials*, *35*(51), p.2301063.

[50] Lee, J.W., Han, J., Kang, B., Hong, Y.J., Lee, S. and Jeon, I., 2025. Strategic development of memristors for neuromorphic systems: low-power and reconfigurable operation. *Advanced Materials*, *37*(19), p.2413916.

[51] Basu, A., Acharya, J., Karnik, T., Liu, H., Li, H., Seo, J.S. and Song, C., 2018. Low-power, adaptive neuromorphic systems: Recent progress and future directions. *IEEE Journal on Emerging and Selected Topics in Circuits and Systems*, *8*(1), pp.6-27.

[52] Barchi, F., Urgese, G., Siino, A., Di Cataldo, S., Macii, E. and Acquaviva, A., 2019. Flexible on-line reconfiguration of multi-core neuromorphic platforms. *IEEE Transactions on Emerging Topics in Computing*, *9*(2), pp.915-927.

[53] Kornijcuk, V. and Jeong, D.S., 2019. Recent Progress in Real-Time Adaptable Digital Neuromorphic Hardware. *Advanced Intelligent Systems*, *1*(6), p.1900030.

[54] Filatov, A., Pogorelov, A. and Pogoryelov, Y., 2014. Negative differential resistance in magnetic tunnel junction systems. *physica status solidi (b)*, *251*(1), pp.172-177.

[55] Khan, A., Sohail, A., Zahoora, U. and Qureshi, A.S., 2020. A survey of the recent architectures of deep convolutional neural networks. *Artificial intelligence review*, *53*(8), pp.5455-5516.

[56] Navas-Olive, A., Amaducci, R., Jurado-Parras, M.T., Sebastian, E.R. and de la Prida, L.M., 2022. Deep learning-based feature extraction for prediction and interpretation of sharp-wave ripples in the rodent hippocampus. *Elife*, *11*, p.e77772.

[57] Abdolrasol, M.G., Hussain, S.S., Ustun, T.S., Sarker, M.R., Hannan, M.A., Mohamed, R., Ali, J.A., Mekhilef, S. and Milad, A., 2021. Artificial neural networks based optimization techniques: A review. *Electronics*, *10*(21), p.2689.

[58] Li, Z., Liu, F., Yang, W., Peng, S. and Zhou, J., 2021. A survey of convolutional neural networks: analysis, applications, and prospects. *IEEE transactions on neural networks and learning systems*, *33*(12), pp.6999-7019.

[59] Huang, J.Y., Syu, J.L., Tsou, Y.T., Kuo, S.Y. and Chang, C.R., 2022. In-Memory Computing Architecture for a Convolutional Neural Network Based on Spin Orbit Torque MRAM. *Electronics*, *11*(8), p.1245.

[60] Gu, J., Wang, Z., Kuen, J., Ma, L., Shahroudy, A., Shuai, B., Liu, T., Wang, X., Wang, G., Cai, J. and Chen, T., 2018. Recent advances in convolutional neural networks. *Pattern recognition*, *77*, pp.354-377.

[61] Rodrigues, D.R., Raimondo, E., Puliafito, V., Moukhadder, R., Azzerboni, B., Hamadeh, A., Pirro, P., Carpentieri, M. and Finocchio, G., 2023. Dynamical neural network based on spin transfer nano-oscillators. *IEEE Transactions on Nanotechnology*, *22*, pp.800-805.

[62] Alzubaidi, L., Zhang, J., Humaidi, A.J., Al-Dujaili, A., Duan, Y., Al-Shamma, O., Santamaría, J., Fadhel, M.A., Al-Amidie, M. and Farhan, L., 2021. Review of deep learning: concepts, CNN architectures, challenges, applications, future directions. *Journal of big Data*, *8*(1), p.53.

[63] Raj, P., Dutta, A., Sharma, P., Zaya, M.H., Das, M., Ghosh, T. and Roy, P., 2024, June. CNN model for handwritten digit recognition with improved accuracy and performance using MNIST dataset. In *2024 4th International Conference on Intelligent Technologies (CONIT)* (pp. 1-9). IEEE.

[64] Seng, L.M., Chiang, B.B.C., Salam, Z.A.A., Tan, G.Y. and Chai, H.T., 2021. MNIST handwritten digit recognition with different CNN architectures. *Journal of Applied Technology and Innovation*, *5*(1).

[65] Madhulika, P.S.S. and Sampath, N., 2022, May. An application of normalizer free neural networks on the SVHN dataset. In *2022 International conference on applied artificial intelligence and computing (ICAAIC)* (pp. 238-242). IEEE.

[66] Pradhan, O., Tang, G., Makris, C. and Gudipati, R., 2024, March. Reading and Understanding House Numbers for Delivery Robots Using the "SVHN Dataset". In *2024 IEEE International Conference on Industrial Technology (ICIT)* (pp. 1-7). IEEE.

[67] Vinay, S.B. and Balasubramanian, S., 2023. A comparative study of convolutional neural networks and cybernetic approaches on CIFAR-10 dataset. *International Journal of Machine Learning and Cybernetics (IJMLC)*, *1*(1), pp.1-13.

[68] Kumar, V. and Rajput, A., 2025, March. CIFAR-10 Object Detection Using. In *Proceedings of International Conference on Recent Trends in Computing: ICRTC 2024, Volume 2* (Vol. 2, p. 313). Springer Nature.

[69] Huang, Jui-Ting, Jinyu Li, and Yifan Gong. "An analysis of convolutional neural networks for speech recognition." In *2015 IEEE International Conference on Acoustics, Speech and Signal Processing (ICASSP)*, pp. 4989-4993. IEEE, 2015.

[70] Dhanjal, A.S. and Singh, W., 2024. A comprehensive survey on automatic speech recognition using neural networks. *Multimedia Tools and Applications*, *83*(8), pp.23367-23412.

[71] Hasnaine, Q.R., Wickard, I., Sakr, H.A., Fouda, M.M. and Ashour, A.F., 2025, September. Low-SNR Robust Modulation Classification on the RadioML Dataset with Deep CNNs. In *2025 3rd International Conference on Artificial Intelligence, Blockchain, and Internet of Things (AIBThings)* (pp. 1-6). IEEE.

[72] Milosheski, L., Bertalanič, B., Fortuna, C. and Mohorčič, M., 2025. Radio Signals Recognition with Unsupervised Deep Learning: A Survey. *IEEE access*, *13*, pp.217769-217798.

[73] Ghanatian, H., Benetti, L., Anacleto, P., Böhnert, T., Farkhani, H., Ferreira, R. and Moradi, F., 2023. Spin–orbit torque flash analog-to-digital converter. *Scientific Reports*, *13*(1), p.9416.

[74] Tammina, S., 2019. Transfer learning using vgg-16 with deep convolutional neural network for classifying images. *International Journal of Scientific and Research Publications (IJSRP)*, *9*(10), pp.143-150.

[75] Zhang, R., Zhu, F., Liu, J. and Liu, G., 2019. Depth-wise separable convolutions and multi-level pooling for an efficient spatial CNN-based steganalysis. *IEEE Transactions on Information Forensics and Security*, *15*, pp.1138-1150.

[76] Solovyev, R.A., Vakhrushev, M., Radionov, A., Romanova, I.I., Amerikanov, A.A., Aliev, V. and Shvets, A.A., 2020, April. Deep learning approaches for understanding simple speech commands. In *2020 IEEE 40th international conference on electronics and nanotechnology (ELNANO)* (pp. 688-693). IEEE.

[77] Zhang, H., Zhou, F., Du, H., Wu, Q. and Yuen, C., 2025. Revolution of wireless signal recognition for 6g: Recent advances, challenges and future directions. *IEEE Communications Surveys & Tutorials*.

Supplementary files for:

# Reconfigurable Multistate MRAM Synapses with Vortex STNO based Neurons for Scalable In-Memory Convolutional Neural Networks

**Ravish Kumar Raj[1,3,5], Simon N. Richter[3], Saeed Baghaee Ivriq[2], Oliver Fridorf[3], Darío Fernández-Khatiboun[1], Yasser Rezaeiyan[1], Sonal Shreya[3], Luana Benetti[4], Tim Boehnert[4], Ricardo Ferreira[4], Hooman Farkhani[1], Sonal Shreya[3] and Farshad Moradi[1, 3]***

[1]The Faculty of Engineering, Institute of Mechanical and Electrical Engineering Department, IME, University of Southern Denmark, Odense

[2]Electronic Circuits and Systems, KU Leuven, Arenberg Campus, Kasteelpark Arenberg, Belgium

[3]Department of Electrical and Computer Engineering, Aarhus University, Aarhus, Aarhus N 8200, Denmark

[4]International Iberian Nanotechnology Laboratory (INL), Braga, Portugal

[5]Department of Electronics and Communication Engineering, Indian Institute of Technology, Roorkee, India 247667

*Email: moradi@sdu.dk

## 1. Device characterization with on-chip (internal) field line

### 1.1. Measurement setup and methods

**Measurement setup:** The device was characterized using a manual probe station equipped with DC probes. Electrical biasing and readout were performed using source-meter units (SMUs), while dynamic and RF signals were monitored using a spectrum analyzer connected through a bias-Tee. All instruments were computer (PC)-controlled via a Python-based measurement program, enabling automated field line sweeps and synchronized data acquisition.

**Figure S1** illustrates the complete experimental setup and measurement methods used for the electrical and RF characterization of the single array sample of 8 MRAM and 1 STNO device.

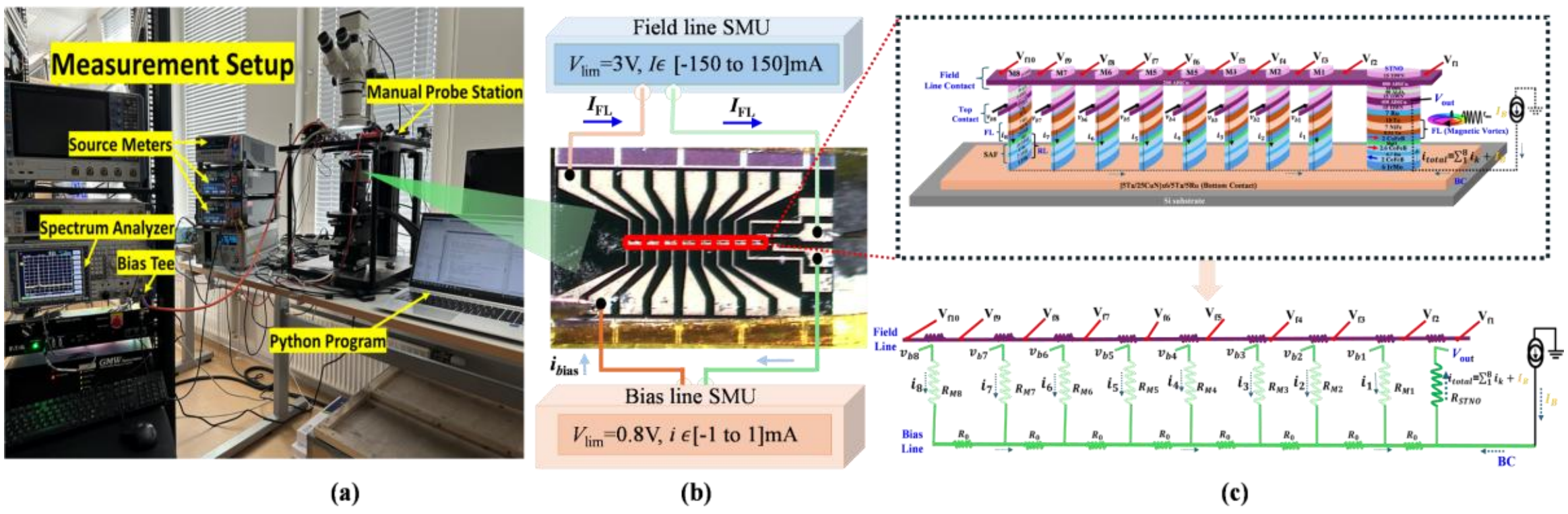

**Figure S1**: Experimental characterization setup and measurement methodology. (a) shows the laboratory setup comprising a manual probe station, source-meter units (SMUs), bias-tee, spectrum analyzer, and a Python-controlled data acquisition system (PC). (b) illustrates the probing configuration and electrical biasing scheme, where the field line is driven by one SMU and the bias line is driven by another one. (c) presents the device structure and layout, showing multiple MTJs connected along a common bottom contact on a Si substrate, together with the corresponding equivalent circuit used for weighted synapse and STNO as neuron.

**MRAM characterization:** As shown in Figure S1 (a), the sample is mounted on the probe-station chuck using thin double-sided tape and contacted using a manual probe station. Four probe terminals are employed: two probes are connected to the field line (FL) and two to the bias line (BL) of the device. A single-array sample fabricated at INL is used for the present measurements. An optical microscope integrated with the probe station is used to accurately align the probes with the device contact pads; the microscope image is streamed to the control computer via OBS. Care is taken to retract the microscope from the probing area before moving the sample to avoid mechanical damage. The magnetic field is generated by driving a current through the field line using an R&S source-measure unit (SMU), thereby controlling the field's strength via the applied current. Electrical biasing of the device is provided by a Keithley 2450 SMU, which is also used to measure the device's resistance to verify proper probe contact.

- **Field line (FL)**: Driven by an SMU with a voltage compliance of $V_{\mathrm{lim}} = 2$ V and current range $I_{\mathrm{FL}} \in [-150, +150]$ mA. This line generates the magnetic field required for device switching.
- **Bias line (BL):** Driven by a separate SMU with a voltage compliance of $V_{\mathrm{lim}} = 0.8$ V and current range $i_{\mathrm{bias}} \in [-1, +1]$ mA, used to bias the magnetic tunnel junctions and reading the resistance state.

**STNO characterization:** For spin-torque nano-oscillator (STNO) measurements, a bias-tee (Figure S1) is used to separate the DC biasing and RF signal paths. The DC current from Keithley SMU is applied to the device through the bias-tee, while the generated RF oscillations are routed to an R&S spectrum analyzer. The spectrum analyzer is used to measure the output power spectral density of the device, with key parameters including frequency span, resolution bandwidth (RBW), and the number of capture points. These parameters are optimized to improve frequency resolution and reduce the noise floor, at the expense of increased acquisition time. Resistance measurements using the Keithley SMU are performed prior to RF measurements to ensure reliable electrical contact.

- **Field line (FL)**: Driven by an SMU with a voltage compliance of $V_{\mathrm{lim}} = 3$ V and current range $I_{\mathrm{FL}} \in [-150, +150]$ mA. This line generates the magnetic field required for device switching.
- **Bias line (BL):** Driven by a separate SMU with a voltage compliance of $V_{\mathrm{lim}} = 1$ V and current range $i_{\mathrm{bias}} \in [-5, +5]$ mA, used to bias the magnetic tunnel junctions and reading the resistance state.

**Device architecture and probing**: Microscopy image of fabricated device that comprises of 8 in-plane elliptical MTJ having dimension (0.1µm x0.5µm) and 1 STNO with circular dimension of 0.35µm x 0.35µm on single array having common and individual field line controlled by Vf1, Vf2, ..Vf10 on top (dotted blue box) for writing path and weighted inputs (bias) from top contact vb1, vb2 …vb8 (dotted green box) to shared bottom contact (BC) pads for reading the states MRAM as a synapse. The MTJ cells are spaced 120 µm apart mostly due to the layout of the contacts pads and to eliminate stray field interactions between the neighboring MTJs. The complete material stack used in the devices investigated in this study consists of (thicknesses in nm [5 Ta / 25 CuN]x6 / 5 Ta / 5 Ru / 6 IrMn / 2.0 CoFe30 / 0.7 Ru / 2.6 $CoFe_{40}B_{20}$ / MgO / 2.0 $CoFe_{40}B_{20}$ / 0.21 Ta / 7 NiFe / 10 Ta / 7 Ru. The MgO barrier was deposited with a wedge profile across the wafer. Probing is done through FL and BL contact nobe with proper rotation

**Equivalent circuit**: An equivalent circuit model is shown at the bottom, where each magnetic tunnel junction is represented as a nonlinear resistive element connected to the shared field line. This model is used to interpret the measured electrical and spectral responses.

### 1.2. Results

**Figure S2** shows the field-line (FL) current sweep hysteresis measurements for MRAM devices 1 through 8 for a single array sample, measured at a constant MRAM bias current of 50 µA. For each device, three consecutive sweeps were performed to evaluate switching behavior, repeatability, and stability. Across MRAM 2-6, the devices exhibit clear and well-defined hysteresis loops in resistance versus FL current, characteristic of robust magnetic switching. The resistance states separate cleanly into high-resistance (antiparallel) and low-resistance (parallel) branches, with abrupt transitions at the switching currents. The close overlap of the three sweeps for each device indicates good cycle-to-cycle repeatability and minimal drift.

The switching currents are generally consistent across the array, with transitions occurring symmetrically in the positive and negative FL current directions (approximately within the ±100-150 mA range, depending on the device). This suggests good uniformity of the free layer coercivity and effective field coupling across the array. Minor variations in switching thresholds and resistance levels are observed from MRAM to MRAM, which can be attributed to expected device-to-device process variations. MRAM 1 shows a more gradual and asymmetric resistance evolution compared to the other devices, with less abrupt switching and a narrower resistance window. This behavior may indicate partial switching, increase thermal effects, or reduce magnetic stability relative to the other cells.

Overall, MRAM 2-6 demonstrate stable, repeatable hysteretic switching, well-separated resistance states, and consistent behavior over multiple sweeps, confirming good functional integrity of the single-array sample under the applied bias and FL current conditions. However, MRAM1, MRAM7 and MRAM8 show little distortion in hysteresis.

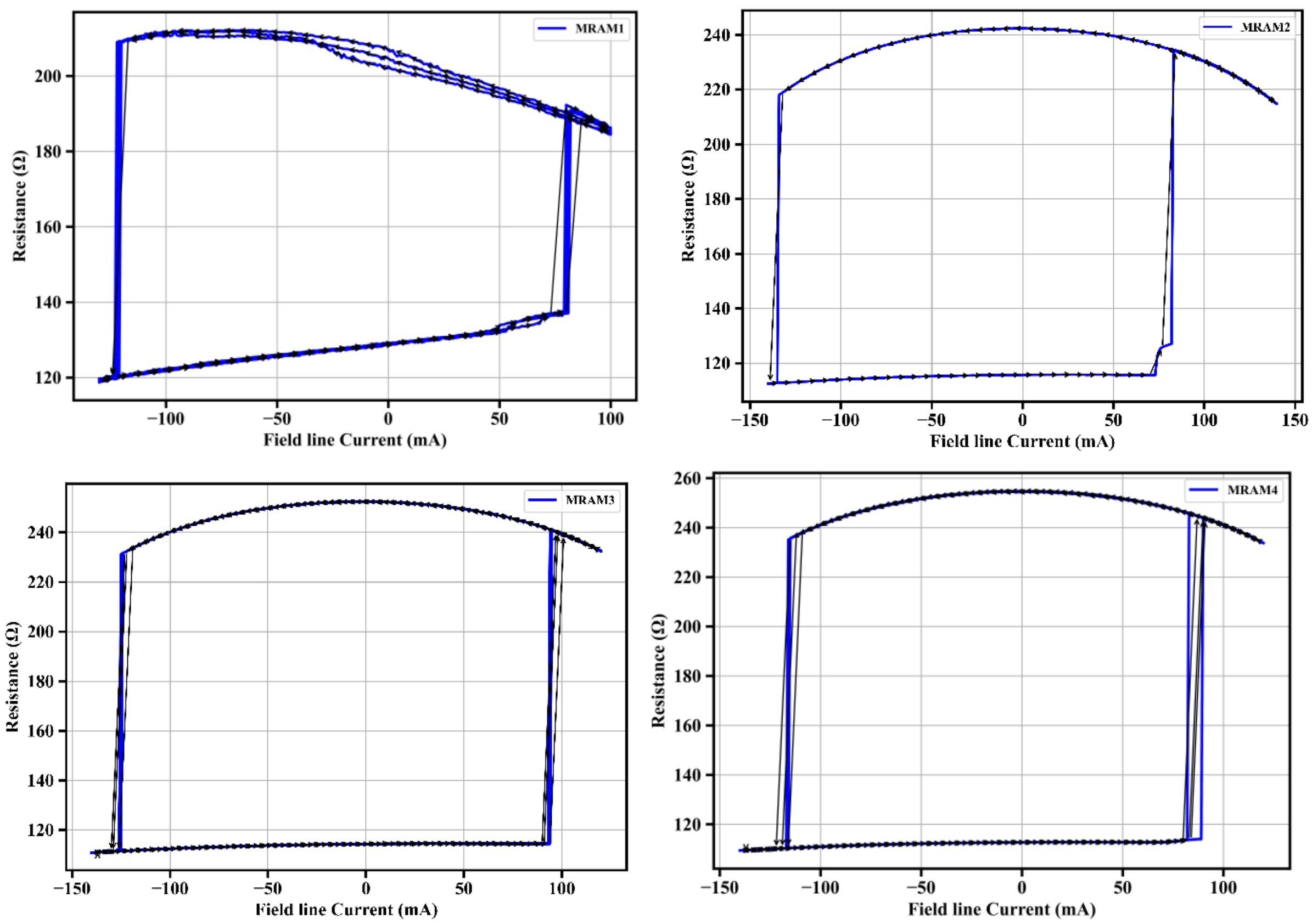

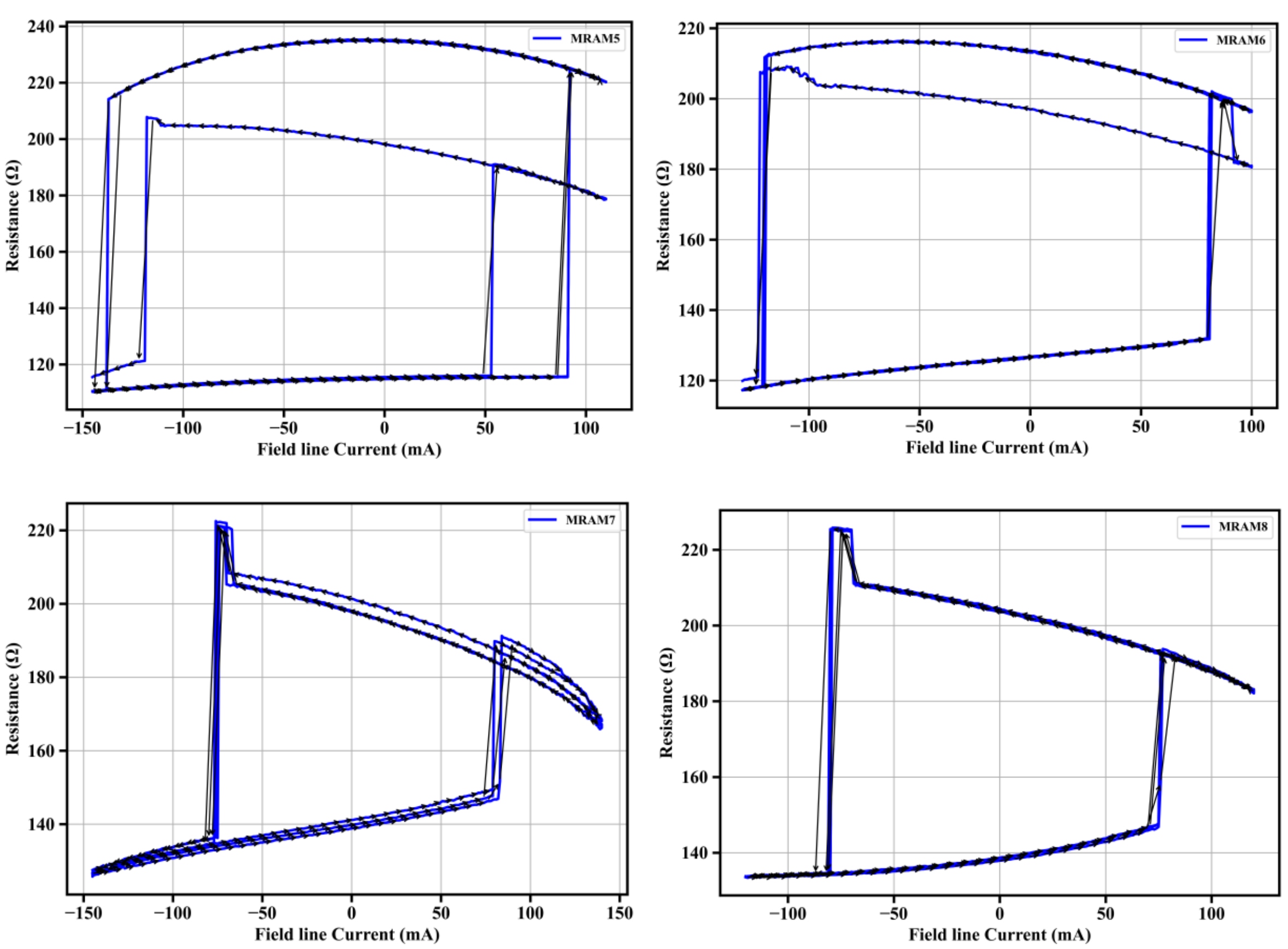

**Figure S2.** On-chip field-line (FL) current-induced hysteresis characteristics of MRAM cells 1-8 within a single array, measured at a constant MRAM bias current of 50 µA. The resistance of each cell is plotted as a function of FL current over three consecutive sweep cycles.

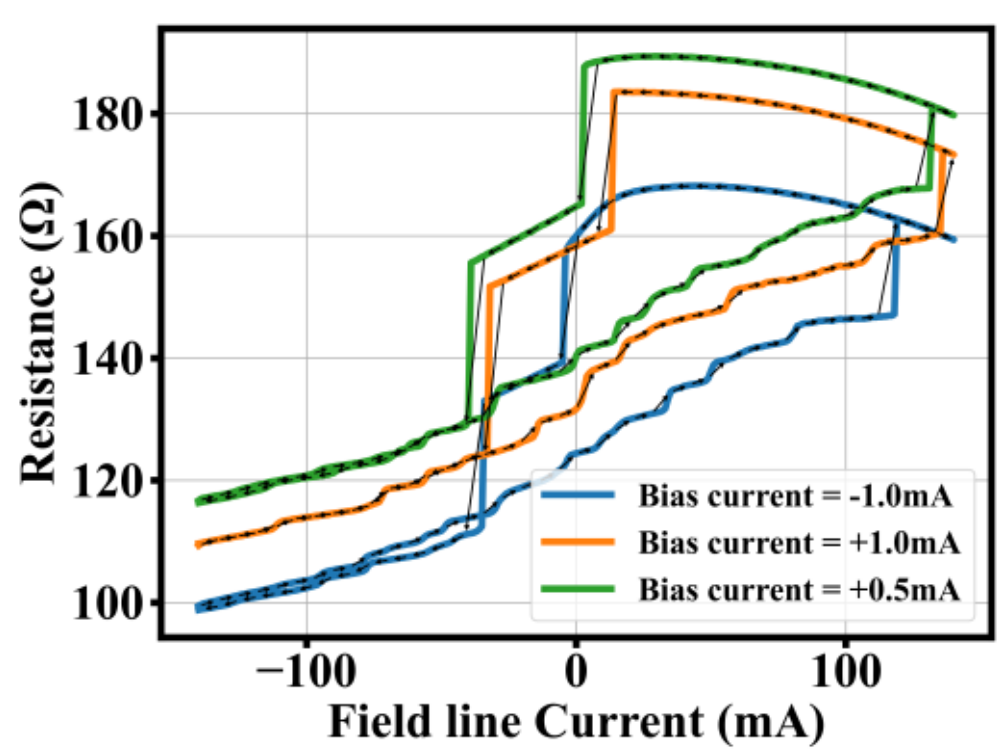

**Figure S3**: Field-line (FL) current sweep hysteresis of a spin-torque nano-oscillator **(STNO)** measured at a constant bias current of 50 µA with FL current swept from −0.14 A to +0.14 A. The device exhibits a multi-state, hysteretic response with an intermediate resistance region attributed to a vortex state

**Figure S3** shows the field-line (FL) current sweep hysteresis of STNO measured on a single array at various bias current, with the FL current swept between −0.14 A and +0.14 A. The STNO exhibits a non-binary, multi-state resistance evolution with pronounced hysteresis, reflecting complex magnetization dynamics rather than abrupt two-state switching. At negative FL currents, the resistance increases gradually with current magnitude, indicating progressive magnetization rotation. A distinct intermediate resistance regime is observed around moderate negative currents, highlighted in the figure and attributed to the formation of a vortex state. As the FL current approaches zero and reverses polarity, a sharp resistance transition occurs, signaling a field-driven reconfiguration of the magnetic state. At positive FL currents, the resistance reaches a higher level and then decreases gradually with increasing current, completing the hysteresis loop. The directional arrows indicate the sweeping direction and emphasize the path-dependent nature of the switching process. The close overlap among the three sweeps demonstrates good repeatability and stability of the STNO magnetic states under repeated FL current cycling. Overall, the measured hysteresis confirms the presence of vortex-mediated switching behavior and highlights the rich magnetic dynamics of the STNO under field-line excitation.

## 2. Device characterization with external field line

### 2.1. Characterization setup and methods

**Figure S4** illustrates the Experimental characterization setup for measurements under an externally applied magnetic field. (a) - (b) Photograph of the device-under-test mounted inside a custom-designed copper toroidal coil used to generate a controllable external magnetic field. The magnetic field strength is tuned by adjusting the coil current supplied to the toroidal structure, allowing systematic investigation of magnetic-field-dependent switching and hysteresis behaviour of the MRAM devices. The probe assembly provides stable electrical contact to the chip during field application. (c) Schematic representation of the electrical biasing configuration and magnetic field distribution. The external field line is driven using an independent SMU to control the magnetic excitation current, while a separate bias line applies the MRAM read/write bias current. The schematic also illustrates the resulting magnetic field profile generated around the device region, highlighting the spatial distribution of the applied field and its interaction with the MRAM cell during characterization.

**MRAM characterization:** Two independent electrical paths are employed. The field line (FL) is driven by an SMU with a voltage compliance of $V_{\mathrm{lim}} = 3$ V and a current range of $\pm 1$ A, generating a local magnetic field along the device through the on-chip field line. The bias line (BL) is driven by a second SMU with a voltage compliance of $V_{\mathrm{lim}} = 0.8$ V and a current range of $\pm 1$ mA. The bias SMU is also used to measure the device resistance prior to each measurement to ensure reliable probe contact. An additional external magnetic field is applied using a copper toroidal coil positioned concentrically around the sample, as shown in Figure S4 (b). A controlled current through the toroid generates a predominantly circumferential magnetic field, with the resulting leakage flux providing a uniform in-plane magnetic field component at the device location. The magnitude and polarity of the external field are controlled by the coil current, enabling systematic field sweeps without mechanical realignment of the sample.

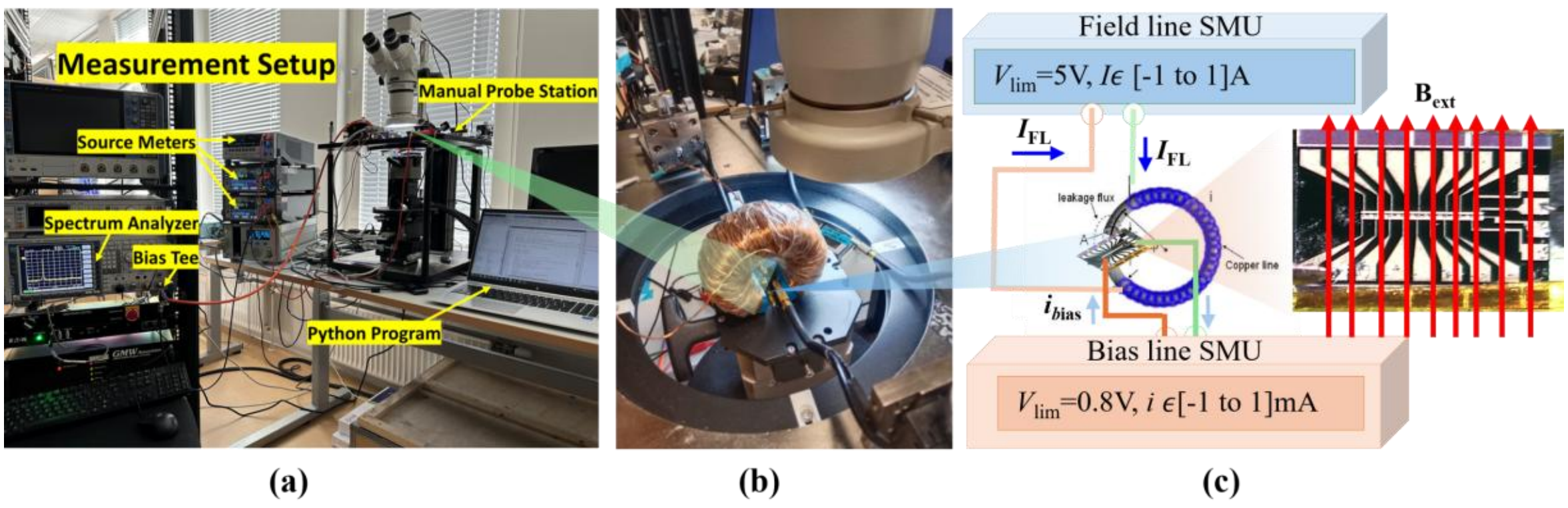

**Figure S4**: Experimental characterization setup with external magnetic field. (a) shows the measurement setup comprising a manual probe station, source-measure units (SMUs), bias-tee, spectrum analyzer, and a Python-controlled data acquisition system. (b) shows the sample positioned within a copper toroidal coil used to generate an external magnetic field via controlled coil current. (c) illustrates the electrical biasing scheme, where the external field line is driven by an SMU and the bias line is driven independently, along with a schematic of the resulting magnetic field distribution across the device.

**STNO characterization:** For spin-torque nano-oscillator (STNO) measurements, a bias-tee is used to separate the DC bias current from the high-frequency voltage oscillations generated by the device. The DC component biases the device through the SMU, while the RF signal is routed to a spectrum analyzer. Measurement parameters such as frequency span, resolution bandwidth (RBW), and acquisition points are optimized to balance spectral resolution, noise floor, and measurement time.

- **Field line** (FL): Driven by an SMU with a voltage compliance of $V_{\text{lim}} = 5$ V and current range $I_{\text{FL}} \in [-1, +1]$ A. This line generates the magnetic field required for device switching.
- **Bias line** (BL): Driven by a separate SMU with a voltage compliance of $V_{\text{lim}} = 1$ V and current range $i_{\text{bias}} \in [-5, +5]$ mA, used to bias the magnetic tunnel junctions and reading the resistance state.

### 2.2. Results

**Figure S5** presents the external field-line (FL) current sweep for external magnetic field generation for hysteresis measurements of MRAM cells 1 to 8 of a single array sample, characterized in the presence of an external magnetic field (applied using an external magnet, toroid). The FL current was swept between −1.0 A and +1.0 A, while the MRAM devices were biased at 50 µA. For each device, three consecutive sweeps were recorded to assess switching behavior, repeatability, and field-induced stability.

Most devices (MRAM 1-8) exhibit sharp, rectangular hysteresis loops with well-defined transitions between low-resistance (parallel) and high-resistance (antiparallel) states. The switching events occur abruptly at relatively high FL currents, reflecting the enhanced magnetic stability and increased switching thresholds resulting from the applied external magnetic field. The close overlap among the three sweeps for each cell demonstrates excellent cycle-to-cycle repeatability and minimal resistance drift. Device-to-device variations are observed in the exact switching current values and resistance levels, which are attributed to local magnetic field nonuniformity, lithographic variations, and differences in magnetic anisotropy across the array. MRAM 8, measured with an opposite sweeper polarity, confirms the reversible and symmetric switching behavior under reversed FL current direction in the presence of the external field. Overall, the measurements demonstrate that the application of an external magnetic field leads to robust, stable, and highly repeatable binary switching in most cells of the single array, while also highlighting isolated non-ideal behavior that provides insight into field–device interactions and array-level variability.

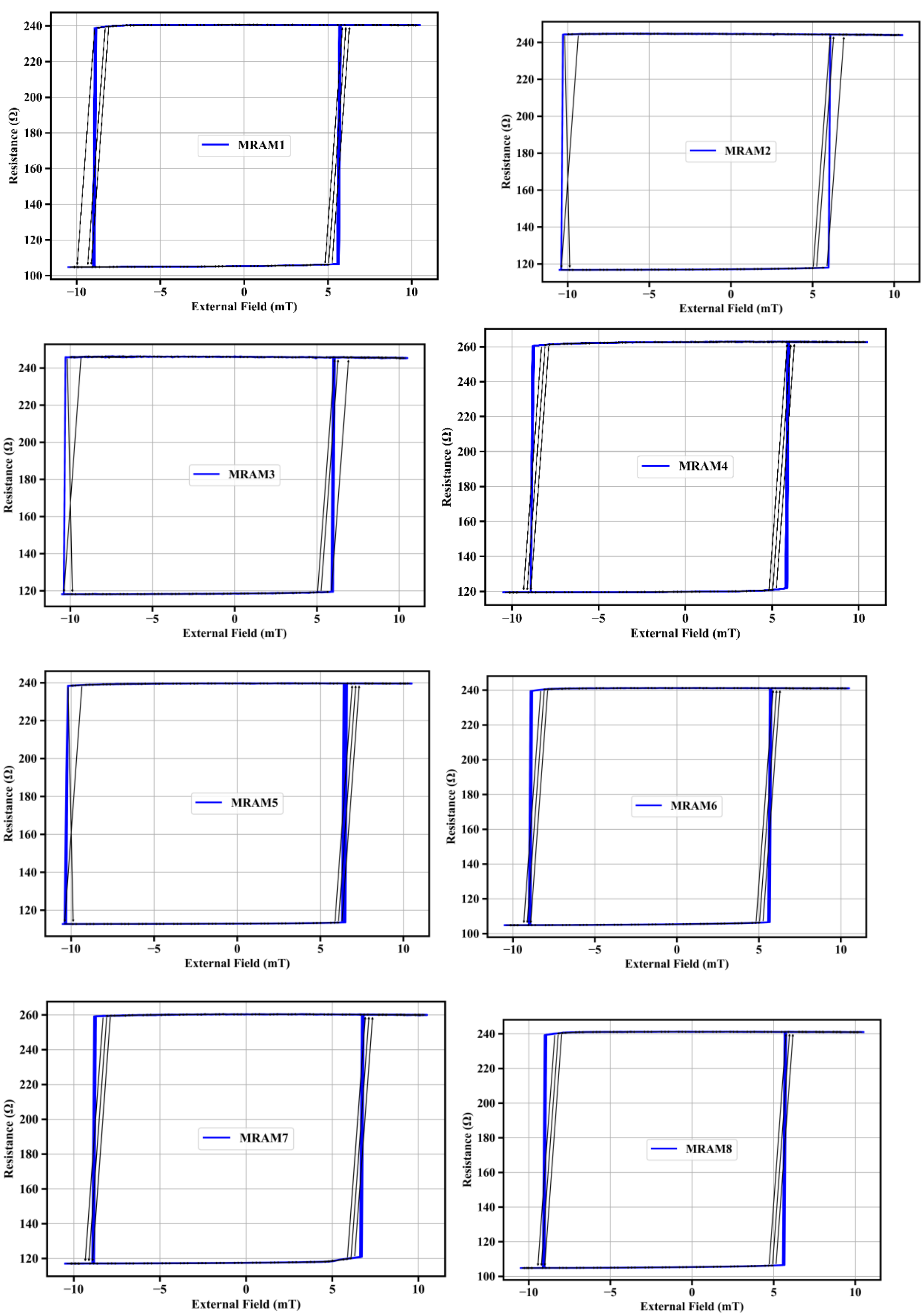

**Figure S5:** External field-line (FL) current-induced hysteresis characteristics of MRAM cells 1-8 from a single array measured in the presence of an external magnetic field. The FL current is swept from -1.0 A to +1.0 A while the MRAM devices are biased at 50 μA and resistance is plotted versus FL current for three consecutive sweeps.